\documentclass[twocolumn,prd,reprint,preprintnumbers,amsmath,amssymb,nofootinbib]{revtex4}
\usepackage{aas_macros}
\pdfoutput=1

\usepackage{epsfig,psfrag,graphics,verbatim}
\usepackage{graphicx}
\usepackage{dcolumn}
\usepackage{bm}
\usepackage{xcolor}
\usepackage{float}
\usepackage{amsmath}
\usepackage{amssymb}
\usepackage{siunitx}
\usepackage{hyperref}
\usepackage{xfrac}

\hypersetup{
  colorlinks=true,
  linkcolor=blue,
  citecolor=blue,
  urlcolor=blue,
  hyperfootnotes=true
}

\newcommand{\be}{\begin{equation}} 
\newcommand{\ee}{\end{equation}}
\newcommand{\bea}{\begin{equation}\begin{aligned}}
\newcommand{\eea}{\end{aligned}\end{equation}}

\newcommand{\Msun}{\text{M}_\odot}
\newcommand\Mcut{M_{\text{cut}}}
\newcommand{\Mhalf}{M_{\sfrac{1}{2}}}
\newcommand\Mhm{M_{\text{hm}}}
\newcommand\khm{k_{\text{hm}}}
\newcommand\Tkd{T_{\text{kd}}}
\newcommand\WDM{\text{WDM}}
\newcommand\FDM{\text{FDM}}

\begin{document}
\title{Implications of Milky Way Substructures for the Nature of Dark Matter}

\author{Mar\'ia Benito}
\email{mariabenitocst@gmail.com}
\affiliation{National Institute of Chemical Physics and Biophysics, R\"avala 10, Tallinn 10143, Estonia}

\author{Juan Carlos Criado}
\email{jccriadoalamo@kbfi.ee}
\affiliation{National Institute of Chemical Physics and Biophysics, R\"avala 10, Tallinn 10143, Estonia}

\author{Gert H\"utsi}
\email{gert.hutsi@to.ee}
\affiliation{National Institute of Chemical Physics and Biophysics, R\"avala 10, Tallinn 10143, Estonia}

\author{Martti Raidal}
\email{Martti.Raidal@cern.ch}
\affiliation{National Institute of Chemical Physics and Biophysics, R\"avala 10, Tallinn 10143, Estonia}

\author{Hardi Veerm\"ae}
\email{hardi.veermae@cern.ch}
\affiliation{National Institute of Chemical Physics and Biophysics, R\"avala 10, Tallinn 10143, Estonia}

\date{\today}

\begin{abstract}
We study how the indirect observation of dark matter substructures in the Milky Way, using recent stellar stream studies, translates into constraints for different dark matter models. Particularly, we use the measured number of dark subhalos in the mass range $10^7$--$10^9 \Msun$ to constrain modifications of the subhalo mass function compared to the cold dark matter scenario.
We obtain the lower bounds $m_\WDM > \SI{3.2}{keV}$ and $m_\FDM > \SI{5.2e-21}{eV}$ on the warm dark matter and fuzzy dark matter particle mass, respectively. When dark matter is coupled to a dark radiation bath, we find that kinetic decoupling must take place at temperatures higher than $\Tkd > \SI{0.7}{keV}$. We also discuss  future prospects of stellar stream observations.
\end{abstract}

\maketitle

\section{Introduction}

Unveiling the elusive nature of dark matter (DM) has proven to be an extremely difficult endeavor~\cite{Jungman:1995df,Bertone:2004pz}. The DM candidate particle masses span more than 75 orders of magnitude while its feeble non-gravitational interactions with the Standard Model, if they exist at all, are strongly model dependent and have so far evaded detection. 
In the absence of any reliable signal of direct or indirect detection, the main source of information to constrain the DM properties are cosmological and astrophysical observations.

The cold dark matter (CDM) paradigm has been remarkably successful in explaining the large scale structure of the Universe. Yet there are several discrepancies between observations and CDM only numerical simulations at smaller scales, that is, galactic and subgalactic scales. 
These include the ``missing satellite" problem~\cite{Klypin:1999uc,Moore:1999nt}, 
the ``too-big-to-fail" problem~\cite{BoylanKolchin:2011de,BoylanKolchin:2011dk,Garrison-Kimmel:2014vqa,Papastergis:2014aba,Kaplinghat:2019svz}, 
the core-cusp problem~\cite{Moore:1994yx,Flores:1994gz,Burkert:1995yz,Moore:1999gc,vandenBosch:2000rza,deBlok:2001hbg}, 
the plane of satellites problem~\cite{Pawlowski:2013kpa} 
and the diversity problem~\cite{Oman:2015xda}. 
It is unknown whether these small-scale discrepancies might be alleviated by baryonic physics or if they arise from an inadequacy of the standard paradigm (for recent reviews see {\it e.g.}~\cite{Kuhlen:2012ft,Tulin:2017ara, 2017ARA&A..55..343B}). The latter solution has motivated alternative DM scenarios, such as warm dark matter (WDM), fuzzy dark matter (FDM) and self-interacting dark matter (SIDM) models.

In this respect, the abundance of DM substructure can provide valuable information about the nature of DM. Many DM models predict abundances of low-mass dark subhalos that lie much below the CDM prediction. Such models are WDM~\cite{Bode:2000gq,Colin:2000dn}, FDM~\cite{Hu:2000ke,Schive:2015kza,Marsh:2015xka,Hui:2016ltb,Du:2016zcv} and those SIDM scenarios in which DM interacts with a dark radiation bath~\cite{CyrRacine:2012fz,Vogelsberger:2012ku,Vogelsberger:2015gpr,Huo:2017vef}. Other models, such as primordial black hole (PBH) DM~\cite{Hawking:1971ei,Carr:1974nx,Carr:2016drx}, may, instead, enhance small scale structure due to the PBH induced shot noise~\cite{Murgia:2019duy,Hutsi:2019hlw,Inman:2019wvr}.

Observables such as gravitational lensing~\cite{Vegetti:2009cz, Vegetti:2012mc, Hezaveh:2014aoa, Hezaveh:2016ltk, 2018PhRvD..98j3517D, 2019arXiv190902005B, Hsueh:2019ynk, Alexander:2019puy, 2020MNRAS.491.6077G}, 
the Lyman-$\alpha$ forest~\cite{Croft:1997jf,Croft:2000hs, Viel:2013apy, Seljak:2006bg, 2009JCAP...05..012B, Irsic:2017yje, 2019arXiv191109073P} 
and stellar dynamics~\cite{2002MNRAS.332..915I, Yoon+2011, Carlberg2012, Feldmann:2013hqa, Bovy:2016irg, 2016MNRAS.463..102E, Buschmann:2017ams, 2017arXiv171106267K, 2018JCAP...06..007E, 2018MNRAS.473.2060J, Marsh:2018zyw, Banik:2018pjp, Schive:2019rrw, 2019ApJ...878L..32N, 2019ApJ...880...38B, Banik:2019cza}
can be used to experimentally distinguish between these DM scenarios. The analysis of fluctuations in the stellar density of tidal streams in ref.~\cite{Banik:2019cza}, used to indirectly measure the number of dark subhalos in the Milky Way in the mass range $10^7$--$10^9 \Msun$, can be used to constrain DM models. The case of WDM has been studied in ref.~\cite{Banik:2019smi}. Here, we generalize this analysis to other DM scenarios and discuss future prospects for similar studies. We find that they may improve the bounds from lensing and Lyman-$\alpha$, which we review.

We discuss halo substructure for different DM scenarios in a unified framework. To this end we show that the modification of the subhalo mass function (SHMF), compared to the CDM case, can be approximated by a universal fitting form for a broad range of DM models. Given the theoretical and astrophysical uncertainties it is possible to constrain the scale at which the SHMF is suppressed, but not the specific shape of the suppression factor. This implies that it is not yet possible to observationally discriminate between \emph{modified} DM scenarios using stellar streams. Nevertheless, the current constraints are on the verge of clashing with the modified DM explanation of small scale discrepancies. Therefore, future observations have the potential to test the validity of the CDM paradigm.

The paper is structured as follows. In section~\ref{sec:DMscenarios}, we review the SHMF for different DM models. Section~\ref{sec:Bounds} describes different techniques used to constrain the SHMF. It further includes a summary of current bounds on distinct DM models. In section~\ref{sec:Prospects} we present future prospects and we conclude in section~\ref{sec:Conclusions}. We use natural units $\hbar = c = 1$.

\section{Halo substructure in modified DM scenarios}
\label{sec:DMscenarios}

The SHMF for WDM, FDM and SIDM models is suppressed for small halo masses when compared to the prediction for a CDM Universe. The deviation can be approximately parametrized by
\be\label{eq:mass-function-f}
	\frac{dn_{\rm sub}}{dM}
	= f
	\left[1 + \left(\frac{\Mcut}{M}\right)^{\alpha}\right]^{-\beta}
	\left(\frac{dn_{\rm sub}}{dM}\right)_{\text{CDM}},
\ee
where $\Mcut$, $\alpha$, $\beta$ and $f$ are free parameters that depend on the
DM model under consideration. This fitting form generalizes the one proposed in
refs.~\cite{Schneider+2012,Banik:2019cza} which corresponds to the case
$\alpha = 1$ (see appendix~\ref{sec:AppendixA} for further discussion). The
survival fraction $f$ takes into account baryonic effects, such as tidal
interactions with the central galaxy's potential, that reduce the number of
subhalos with respect to the DM only case.  In the CDM scenario, $f$ is expected
to take values in the range 0.1 - 0.5~\cite{Onguia+2009, Sawala+2017,
  GarrisonKimmel+2017}. Although the effect of baryonic tidal stripping and
disruption is yet to be studied in detail, it can be neglected for the purpose
of obtaining conservative bounds on $\Mcut$, as it would reduce the number of
subhalos with masses around $\Mcut$.  In order to better separate the scale of
suppression and the shape of the suppression factor, we also define the related
mass scale $\Mhalf \equiv \Mcut / \sqrt[\alpha]{2^{1/\beta} - 1}$, which gives
the point at which the abundance of subhalos is suppressed by a factor of $1/2$
with respect to the CDM case. Constraints in the $(f, \Mhalf)$ plane for
different shapes of the suppression factor~\eqref{eq:mass-function-f} are
depicted in figure~\ref{fig:contours}. We remark that
eq.~\eqref{eq:mass-function-f} also contains the exponential suppression factor,
which is obtained in the limit $\beta \to \infty$ with $\Mhalf$ kept fixed.%
\footnote{%
  Note that $\Mcut$ vanishes in this limit.%
}

Another important mass scale related to the suppression of the linear power spectrum of density fluctuations is the half-mode mass
\be\label{eq:mhm}
	\Mhm \equiv \frac{4}{3} \pi \rho_b \left(\frac{\pi}{\khm}\right)^3  ,
\ee
where $\rho_b=2.8\times10^{11}\Omega_m h^2\,{\rm M_{\odot}/Mpc^3}$ is the background matter density and $\khm$ is the wavenumber for which the linear power spectrum is suppressed by a factor of $1/2$ with respect to the CDM case. We stress that $\Mhalf$, $\Mcut$ and $\Mhm$ are different mass scales. The suppression of the SHMF can be related to the modified linear spectrum via the adimensional parameter $\gamma = \Mcut / \Mhm$.

The simplified ansatz \eqref{eq:mass-function-f} can fail in the regime where the SHMF is strongly suppressed with respect to the CDM case. As shown in appendix~\ref{sec:AppendixA} a better fit can be obtained by allowing for a weak dependence in the subhalo mass. In addition, $f$ might not be constant in the mass range under study ({\it i.e.} $10^{7}-10^{9}\,{\rm M_{\odot}}$) \cite{Onguia+2009} and can depend on the DM model under consideration since subhalos with shallower cores are more prone to depletion~\cite{Errani&Penarrubia2019}.
However, as the constraints on the abundance of subhalos are determined mainly by the suppression around the scale $\Mhalf$, modifications to the SHMF much below this mass scale, {\it e.g.} additional peaks as seen in FDM studies~\cite{Du:2016zcv}, do not affect our conclusions.

In the following we will review the SHMF in eq.~\eqref{eq:mass-function-f} for different DM models. 

\subsection*{Warm dark matter}

Warm DM particles posses a non-negligible velocity dispersion. This would suppress primordial matter fluctuations at small scales.
The values $\alpha=1$, $\gamma = 2.7$ and $\beta = 0.99$ have been found in ref.~\cite{Lovell:2013ola} for a thermal WDM relic. However, other values for these two parameters have been suggested~\cite{Schneider+2012, Lovell:2013ola}. For this reason, in appendix~\ref{sec:AppendixA} we further investigate substructure suppression in WFM (and FDM) models using the analytic formalism of~\cite{Schneider:2014rda} which provides a rather generic framework for treating models with reduced small-scale power.

The scale $\Mhm$ is controlled by the mass $m_\WDM$ of the WDM particle through~\cite{Viel+2005}:
\begin{multline}
  \label{eq:khm-wdm}
  M_{\rm hm}
  \simeq
  2\times10^{10} {\rm M_{\odot}}
  \left(\frac{m_\WDM}{\si{keV}}\right)^{-3.33} \\
  \times \left(\frac{\Omega_m}{0.30}\right)
  \left(\frac{\Omega_\WDM}{0.25}\right)^{0.33}
  \left(\frac{h}{0.7}\right)^{2.66},
\end{multline}
where $\Omega_m$ and $\Omega_\WDM$ are the contributions of matter and WDM to the density parameter, respectively, and $h$ is the dimensionless Hubble constant.

\subsection*{Fuzzy dark matter}

FDM is a theoretically well-motivated scenario where DM is constituted by ultralight scalars such as axion-like particles or moduli fields~\cite{Marsh:2015xka,Hui:2016ltb}. In this scenario, halos with masses below some sharp cutoff scale are strongly suppressed due to quantum pressure effects~\cite{Hu:2000ke,Marsh:2016vgj}. The effect on the SHMF can be approximated by eq.~\eqref{eq:mass-function-f} with a relatively high value of $\beta$, with $\alpha=1.1$ and $\beta=2.2$~\cite{Schive:2015kza}. The cutoff scale is dictated in this case by the particle mass $m_\FDM$~\cite{Hui:2016ltb},
\be\label{eq:Mhm-fdm}
  \Mhm
  \simeq
  4.6 \times 10^{10} \Msun
  \left(\frac{\Omega_\FDM}{0.25}\right)
  \left(\frac{\SI{e-22}{eV}}{m_\FDM}\right)^{4/3}.
\ee
In appendix~\ref{sec:AppendixA} we give a slight modification of the above formula (see eq.~\eqref{ec:khm_updated}). Constraints on $m_{\FDM}$ obtained by using eq.~\eqref{ec:khm_updated} differ by less than 5\% compared to those obtained  by using eq.~\eqref{eq:Mhm-fdm}.

We can further use the quantum nature of FDM sub-halos to set constraints on the FDM mass. Particularly, FDM particles inside the solitonic core of a subhalo, that is orbiting a host galaxy, have a finite probability of tunneling the subhalo's self-gravitational potential. Thus, after a finite time the subhalo might be completely disrupted~\cite{Hui:2016ltb}. The characteristic timescale $\tau \equiv M_{\rm sub}/\dot M_{\rm sub}$ for the depletion of a FDM subhalo, with mass $M_{\rm sub}$ and solitonic central density $\rho_c$, can be expressed as $\tau = T_{\rm orb} C^{-1}(\rho_c/\bar{\rho}_{\rm host})$ where $\bar{\rho}_{\rm host}$ is the average density of the host within the orbital radius, $T_{\rm orb}$ is the orbital period and $C$ is an invertible function determined from the Schr\"odinger-Poisson system for FDM halo~\cite{Hui:2016ltb,Du:2018zrg}. Therefore, FDM subhalos with a solitonic central density 
\be
	\rho_c < C(N_{\rm orb})\bar{\rho}_{\rm host}
\ee
have lost a significant amount of its mass after having completed more than $N_{\rm orb}$ circular orbits around its host halo. We approximate $C(N_{\rm orb})$ with the analytic fitting form given in ref.~\cite{Du:2018zrg}.

The central density of the solitonic core satisfies $\rho_c \leq 0.0044 (G m_{\rm FDM}^2)^3 M^4$~\cite{Hui:2016ltb}. Therefore, observations of FDM subhalos with a mass  $M_{\rm sub}$ imply the following constraint for the FDM mass:
\begin{multline}\label{eq:mFDM_tidal}
    m_\FDM \geq
    \SI{e-22}{eV}
    \left[\frac{M_{\rm sub}}{10^9\,{\rm \Msun}}\right]^{-2/3}
    \left[\frac{\bar{\rho}_{\text{host}}(r)}{7.05\,{\rm \Msun}\si{pc^{-3}}}\right]^{1/6}
    \\ \times
    \left[C(N_{\rm orb})\right]^{1/6},
\end{multline}
with $\bar{\rho}_{\rm host}(r)$ the averaged density of the host galaxy within radius $r$.\footnote{The constraint can be expressed in terms of the orbital period via $\bar{\rho}_{\rm host} = T_{\rm orb}^{-2} 3\pi/G$ and $N_{\rm orb} = t/T_{\rm orb}$, with $t$ the age of the subhalo.}

The fitting form we use for $C(N_{\rm orb})$ applies for spherical subhalos on circular orbits and further assumes spherical hosts. These assumptions yield a more conservative disruption rate because halos that pass closer to the galactic center have shorter lifetimes.
Non-sphericity of DM substructures has been considered in ref.~\cite{Alexander:2019qsh}.

\subsection*{Self-interacting dark matter}

For SIDM models in which DM couples to a dark radiation bath in the early universe, the elastic scattering between the two species damps the linear power spectrum until the species kinetically decouple at temperature $\Tkd$~\cite{Boehm:2001hm,Buckley:2014hja,Boehm:2014vja}. Numerical simulations show that this leads to a SHMF of the form shown in eq.~\eqref{eq:mass-function-f} with $\alpha=1$ and $\beta = 1.34$ and the suppression scale of the halo mass function~\cite{Vogelsberger:2015gpr,Huo:2017vef}
\be\label{eq:Mcut-sidm}
	\Mcut \simeq 7 \times 10^7 \Msun \left(\frac{\Tkd}{\rm keV}\right)^{-3},
\ee
which corresponds roughly to the horizon mass at kinetic decoupling. This estimate for the suppression of the SHMF was extrapolated from the halo mass function and thus it does not account for the disruption of the halo substructure which may be enhanced in comparison with CDM due to the cored density profiles of DM subhalos~\cite{Vogelsberger:2012ku,Vogelsberger:2014pda}.

The situation can be more involved for atomic DM scenarios~\cite{Kaplan:2009de}, where DM consists of hydrogen like states of dark fermions bound by a light dark mediator. In this case, kinetic decoupling of dark plasma from the dark radiation bath can be a fairly sudden event caused by a dark recombination. This leaves an imprint on density fluctuations at the scale  of the dark-plasma sound horizon, {\it i.e.} the scale $r_{\rm DAO}$ of dark acoustic oscillations (DAO), which is generically much smaller than the scale of baryonic acoustic oscillations. As sub-horizon perturbations are strongly damped before dark recombination, atomic DM predicts a sharp lower bound for the halo mass $M_{\rm min} = 4\pi/3 r_{\rm DAO}^{3} \rho_b$~\cite{CyrRacine:2012fz}.

In SIDM  models, the DM small scale structure can be modified not only via to the suppression of the linear power spectrum, but also due to the modified dynamics of DM halos. However, numerical simulations of a range of SIDM models show that, although DM self-interactions can induce cored density profiles of subhalos, the subhalo abundance of Milky Way like galaxies is unaffected for allowed DM self-interaction cross-sections~\cite{Vogelsberger:2012ku,Vogelsberger:2015gpr}.

\begin{figure*}
  \centering
  \includegraphics[width=\textwidth]{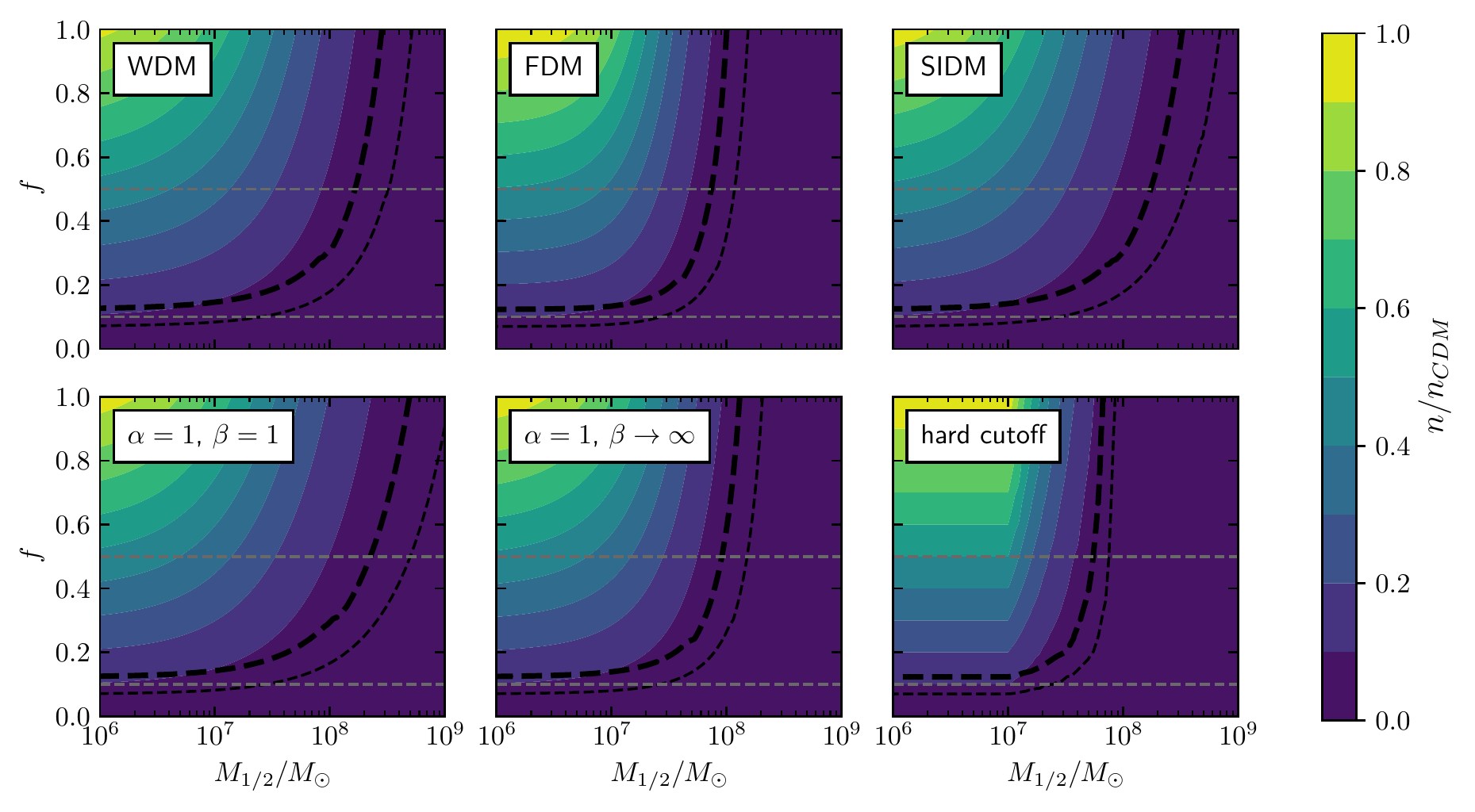}
  \caption{Number of sub-halos $n_{\rm sub}$ in the mass range $10^7$--$10^8 \Msun$ relative to the CDM prediction $n_{\text{CDM}}$, as a function of the parameters $f$ and $M_{1/2}$ for different models. The parameters for named models are specified in section \ref{sec:DMscenarios}. The gray dashed lines show the values between which $f$ is found to be in CDM simulations. The thick and thin black dashed lines show the $68\%$ and $95\%$ CL exclusion limits, respectively, using the analytical likelihood given in ref.~\cite{Banik:2019cza} for the number of halos in the mass bins $10^7$--$10^8 \Msun$ and $10^8$--$10^9 \Msun$. We assume that the log-likelihood follows a $\chi^2$ distribution with 2 degrees of freedom. A hard cutoff corresponds to a SHMF described by a step function such as below $M_{1/2}$, the number of subhalos is zero.
 }
  \label{fig:contours}
\end{figure*}

\section{Bounds on small-scale clumpiness of the matter distribution and implications for DM phenomenology}
\label{sec:Bounds}

Several methods have been proposed and used to place constraints on small-scale  DM distribution. Below we give a brief summary of some of the most common techniques.

\subsection*{Stellar dynamics bounds}
\label{sec:streams}

The presence of dark substructures can be inferred from the dynamical perturbations they induce in their host stellar systems. However, these signals have to be disentangled from the influences of the `usual astrophysics' like globular clusters, molecular clouds, galactic bars, etc. Naturally, the demand for the existence of detailed kinematical data limits the focus here on our own Galaxy.

Several particular probes have been proposed but the full power of this broad direction has still to be realized. For example in~\cite{Feldmann:2013hqa} the authors estimate that by studying the perturbations in the Galactic disk it should be possible to infer the existence of dark substructures with masses $\sim 10^8-10^9\,\Msun$. For masses above $10^9 \Msun$, the SHMF is constrained by the abundance of satellite galaxies in the MW~\cite{2018MNRAS.473.2060J, 2017arXiv171106267K, 2018JCAP...06..007E, 2019ApJ...878L..32N}. In~\cite{Buschmann:2017ams} a method that searches for characteristic wakes left by the passing substructures in the stellar kinematics of the halo stars is shown to provide sensitivity down to subhalo masses $\sim 10^7\,\Msun$. The main advantage of using halo stars far away from the galactic disk is the reduced contamination from astrophysical backgrounds. However, the hotter the stellar system used for probing substructures, the weaker the bounds one expects to obtain.

The central star cluster in the ultra-faint dwarf galaxy Eridanus II can provide a test for FDM models. In ref.~\cite{Marsh:2018zyw} it was argued that this cluster is disrupted by the oscillations of the solitonic core if the FDM particle mass is in the range $10^{-21}$--$\SI{e-19}{eV}$. However,  it was shown in ref.~\cite{Schive:2019rrw} that this effect disappears if tidal stripping produced by the MW potential is taken into account.

One of the most promising bounds on dark subhalo population at the moment can be deduced from the detailed observations of surface density fluctuations in stellar streams \cite{2002MNRAS.332..915I, Yoon+2011, Carlberg2012, 2016MNRAS.463..102E, Bovy:2016irg, Banik:2018pjp, Banik:2019cza, 2019ApJ...880...38B}
(for a review on stellar streams as probes for the nature of DM see~\cite{2016ASSL..420..169J}.)
This method might be sensitive to substructure masses as low as $\sim 10^5 \Msun$~\cite{Yoon+2011, Carlberg2012, Bovy:2016irg}.

The Milky Way subhalo abundance in the mass ranges $10^7$--$10^8 \Msun$ and $10^8$--$10^9 \Msun$ has been estimated from density fluctuations in the density of the cold stellar streams Palomar 5 and GD-1~\cite{Banik:2019cza}. The linear density power spectrum of these streams has been found to be in good agreement with the subhalo abundance predicted by CDM and, thus, in tension with being generated by baryonic structures only.

By fitting the SHMF in eq.~\eqref{eq:mass-function-f} to the subhalo abundance
in the mass range $10^7-10^9 \Msun$ given in~\cite{Banik:2019cza}, we obtain a
rough estimate of the cutoff scale $\Mhm$ for different DM scenarios. The
predicted number of dark subhalos at a given mass range and DM model can be
obtained by integrating the SHMF in eq.~\eqref{eq:mass-function-f}. By comparing
predicted and measured subhalo abundances in the mass range $10^7-10^9 \Msun$ we
set constraints in the plane $(f, M_{1/2})$ for the different DM scenarios
discussed in section~\ref{sec:DMscenarios} and in some idealized toy models (see
figure~\ref{fig:contours}). For this purpose, we build a likelihood function
which is given by the sum of the analytical probability distribution functions
given in~\cite{Banik:2019cza} (cf. equations 6-8) for the subhalo abundance at
mass ranges $10^7-10^8$ and $10^8-10^9 \Msun$. The resulted log-likelihood is
assumed to follow a $\chi^2$ with two degrees of freedom. Thus, we compute the
68\% and 95\% CL exclusion regions which are given by the thick and thin black
dashed lines, respectively, in figure~\ref{fig:contours}.  We remark that, at
higher confidence levels, the suppression of the SHMF can not be constrained
using the numerical likelihood given in ref.~\cite{Banik:2019cza} since current
observations are not sensible to a low number of subhalos at the given mass
range. Moreover, in this case the analytic likelihood fails to approximate the
tails of the numerical one.

We find a relatively mild dependence on the shape of the suppression factor specified by parameters $\alpha$ and $\beta$, with the extremal case of a hard cut-off providing the most stringent constraints. In particular, the largest uncertainties are seen to be contained in the survival factor $f$.

Using a log-uniform prior for $f$ in the range $f \in [0.001, 0.5]$ the upper bound $\Mhm < 4.3 \times 10^7 \Msun$ was reported in ref.~\cite{Banik:2019cza} at 95\% CL for the case of a WDM candidate.  According to eq.~\eqref{eq:khm-wdm}, this is equivalent to the 95\% CL lower limit $m_\WDM > \SI{6.3}{keV}$~\cite{Banik:2019cza}. Eq.~\eqref{eq:Mcut-sidm} translates this bound into into the constraint $T_{\text{kd}} > \SI{1.2}{keV}$ for the SIDM kinetic decoupling temperature. Similarly, eq.~\eqref{eq:Mhm-fdm} implies a lower bound $m_\FDM > \SI{1.9e-20}{eV}$ for the mass of the FDM particle. However, since the SHMF depends on the full shape of the power spectrum and on other details of the DM model, it is not possible to make accurate inferences by simply equating the $\Mhm$ of different models.

In the range $f = 0.1 - 0.5$, shown by the dashed horizontal lines in figure~\ref{fig:contours}, the 95\% CL exclusion boundaries implied for the models under consideration can be summarized as follows:
  \begin{itemize}
  \item WDM:
    \begin{align}
      \Mhm &< [3.4 - 43] \times 10^7 \Msun, \\
      m_\WDM &> [3.2 - 6.8] \, \si{keV}.
    \end{align}
  \item FDM:
    \begin{align}
      \Mhm &< [5.4 - 24] \times 10^7 \Msun, \\
      m_\FDM &> [5.2 - 16] \times 10^{-21} \, \si{eV}.
    \end{align}
  \item SIDM:
    \begin{align}
      \Mcut &< [1.9 - 25] \times 10^7 \Msun, \label{eq:SIDM-bounds-M} \\
      \Tkd &> [0.66 - 1.5] \, \si{keV}. \label{eq:SIDM-bounds-T}
    \end{align}
  \end{itemize}
The weaker bounds correspond to $f=0.5$ and the stronger ones to $f=0.1$. The WDM constraint $m_\WDM > \SI{6.3}{keV}$ in ref.~\cite{Banik:2019cza} corresponds to $f \simeq 0.11$. The SIDM bounds are obtained using the suppression factor in eq.~\eqref{eq:mass-function-f} with $\alpha = 1$ and $\beta = 1.34$. Slightly stronger bounds are found if an exponential suppression factor is used instead.

Stellar kinematical data of MW dwarf galaxies~\cite{Marsh&Pop2015, GonzalezMorales+17, Broadhurst+19, Wasserman+19} favors cored DM profiles than can be explained within the FDM scenario for $m_\FDM < 10^{-21}\,{\rm eV}$. It is important to highlight that the difference between our bound and these limits may reflect systematic errors on the dynamics of dwarf galaxies~\cite{Oman+16a} rather than a tension in the FDM model. In fact, our lower bound on $m_{\FDM}$ is perfectly compatible with limits set by other experimental strategies, such as Lyman-$\alpha$~\cite{2017PhRvL.119c1302I, 2017PhRvD..96l3514K}, CMB \cite{2018MNRAS.476.3063H} or X-ray observations~\cite{2020PhRvD.101b3508M}.

\begin{figure}
  \centering
  \includegraphics[width=\linewidth]{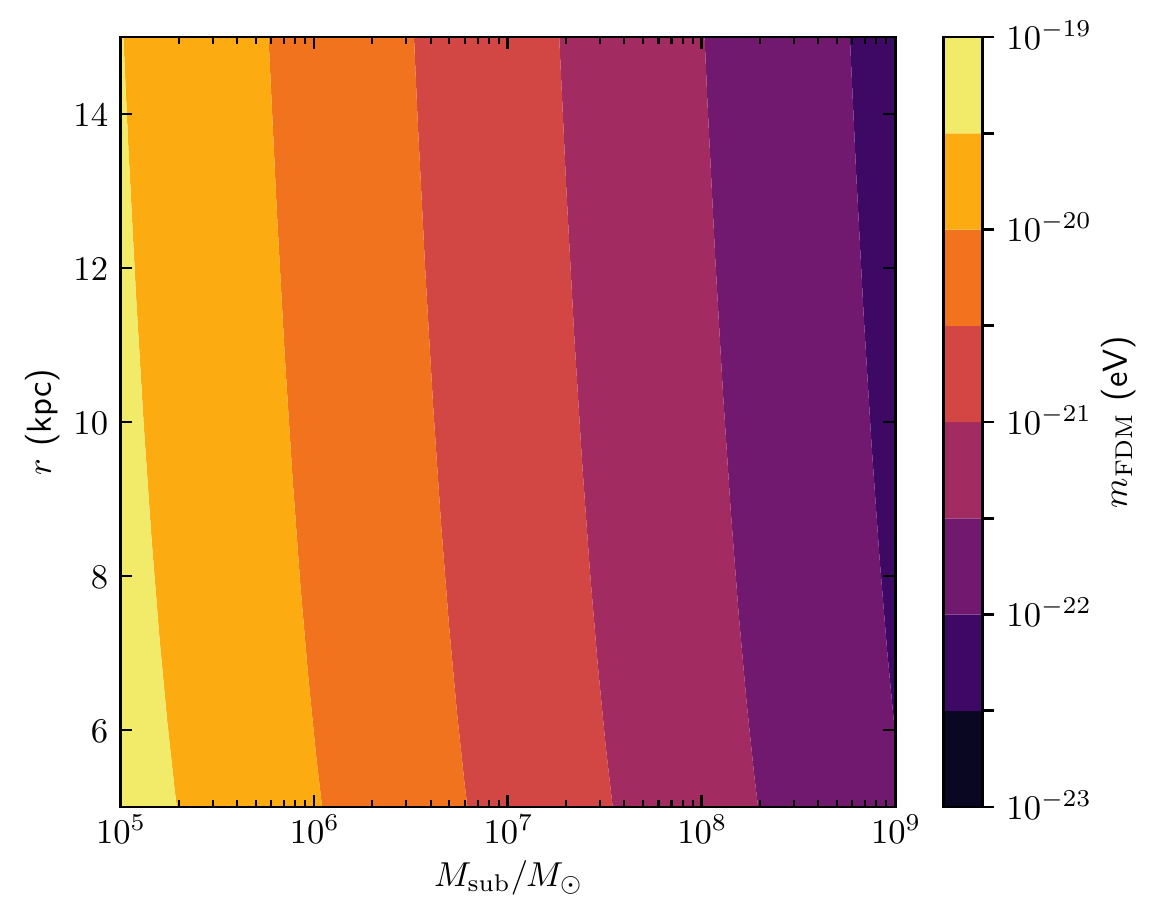}
  \caption{Exclusion boundary for $m_\FDM$ assuming subhalos of mass $M_{\rm sub}$ are observed at distance $r$ to the center of the Milky Way, as a function of both $M_{\rm sub}$ and $r$.}
  \label{fig:m-FDM-contours}
\end{figure}

Figure~\ref{fig:m-FDM-contours} shows the bounds on $m_\FDM$ obtained from eq.~\eqref{eq:mFDM_tidal}. We use $t_{\rm sub}=t_{\rm MW}=13.4\,{\rm Gyr}$ \cite{2004A&A...426..651P} for the age of the subhalos, thus obtaining conservative bounds. We further adopt the linear parametrization of $V(r)(=2\pi r/T_{\rm orb}(r))$ for the Milky Way~\cite{Eilers+2019}:
\be
	V(r) = \SI{229}{\km\,\second^{-1}} - (\SI{1.7}{\km\,\second^{-1} kpc^{-1}}) (r - R_0).
\ee
where $R_0 = 8.122\pm0.031\,{\rm kpc}$ is the distance from the Sun to the center of the Galaxy~\cite{Abuter:2018drb}. The existence of dark subhalos with masses of $10^{7}\,\rm M_{\odot}$ at the perigalacticon of the GD-1 stream $r \simeq \SI{14}{kpc}$, would then imply a FDM mass satisfying
\be
	m_{\rm FDM} \gtrsim10^{-21}\,{\rm eV}.
\ee

Constraints inferred using density fluctuations in cold stellar streams might be subject to different sources of systematic error. For instance, it has been recently pointed out that the GD-1 stream, used in the analysis of \cite{Banik:2019cza}, might have been perturbed by the Sagittarius dwarf galaxy \cite{2020arXiv200107215B}. Thus, in order to set strong constraints on the DM particle nature, it is important to complement results from different experimental strategies.

\subsection*{Lyman-$\alpha$ bounds}

For almost two decades the most stringent and reliable bound on small-scale matter clustering has been obtained via the statistical study of quasar Ly-$\alpha$ forest data, which probes clustering at mildly nonlinear regimes in the redshift interval $z\sim 2-6$, see {\it e.g.}~\cite{Croft:1997jf,Croft:2000hs,Seljak:2006bg, 2009JCAP...05..012B}. For instance, the matter power spectrum $P(k)$ at comoving wavenumber $k\simeq 10\,h{\rm Mpc}^{-1}$ cannot be suppressed more than $10\%$ with respect to the standard $\Lambda$CDM case~\cite{Viel:2013apy}. This has allowed to set a mass limit on thermal relic WDM particle $m_{\rm WDM}>3.3$~keV at $2\sigma$ CL and corresponds to an effective mass-scale suppression in the halo mass function at $\sim 2\times 10^8\,h^{-1}\Msun$~\cite{Viel:2013apy}. The above WDM bound translates to the FDM particle mass limit $m_{\rm FDM}\gtrsim 2\times 10^{-21}$~eV, consistent with the mass bounds found in~\cite{Irsic:2017yje}.

The most recent Ly-$\alpha$ bound is even stronger, ruling out WDM masses below $10$~keV at $95\%$ CL.~\cite{2019arXiv191109073P}. Using eqs.~\eqref{eq:khm-wdm}, \eqref{eq:Mcut-sidm} and~\eqref{eq:Mhm-fdm}, this bound roughly translates into $\Tkd > \SI{2.0}{keV}$ and $m_\FDM > \SI{5.9e-20}{eV}$. However, considering realistic astrophysical uncertainties on the modeling of the intergalactic medium, this result is highly controversial~\cite{Garzilli:2019qki}.

SIDM scenarios have been considered in the context of Ly-$\alpha$ observations in ref.~\cite{Archidiacono:2019wdp} and Ly-$\alpha$ constraints on the PBH abundance have been derived in~\cite{Murgia:2019duy}.

\subsection*{Lensing bounds}

More recently a direct detection of substructures in high resolution imaging data of giant strong lensing arcs has become a new competitive method, currently enabling to detect subhaloes with masses down to $\sim 10^8-10^9\,\Msun$. For example in~\cite{Vegetti:2009cz} a Hubble Space Telescope imaging has been used to detect substructure with mass $\sim 3.5\times 10^9\,\Msun$ in a lens galaxy at redshift $z\sim 0.22$ having total DM mass $\sim 3.6\times 10^{11}\,\Msun$ inside the effective radius. In~\cite{Vegetti:2012mc} Keck adaptive optics imaging has revealed a substructure with mass $\sim 1.9\times 10^8\,\Msun$ in a Sagittarius-size lens galaxy at redshift $z\sim 0.88$. In~\cite{Hezaveh:2016ltk} a subclump with mass $\sim 9\times 10^8\,\Msun$ has been detected using ALMA observations of the strong lensing system SDP.81.

Instead of the above direct subclump detection these techniques can be extended via statistical fluctuation analyses to probe the presence of substructures below a direct detection limit, potentially allowing to reach down to masses $\sim 10^{7}\,\Msun$, see {\it e.g.}~\cite{Hezaveh:2014aoa, 2018PhRvD..98j3517D, 2019arXiv190902005B}.

Yet another approach to constrain substructures relies on measurements of flux ratios of strongly lensed quasars, which allows to indirectly probe the presence of subclumps with masses $\sim 10^6-10^9\,\Msun$~\cite{Hsueh:2019ynk}. This has allowed to constrain the thermal relic WDM particle mass $m_{\rm WDM}>3.8$~keV at $2\sigma$ CL~\cite{Hsueh:2019ynk}, {\it i.e.}, comparable to the level of the above Ly-$\alpha$ bounds. Since these constraints rely on the non-linear substructure, they cannot be straightforwardly mapped to bounds on other DM models via $\Mhm$ comparisons.

\vspace{10pt}

\section{Prospects}
\label{sec:Prospects}

A subhalo interaction with a tidal stream perturbs the energy distribution for the member stars of the latter. This perturbation in orbital energy evolves into fluctuations in surface density at different angular scales depending on the mass and velocity of the perturber, the impact parameter and look-back time at which the interaction took place \cite{Yoon+2011, Carlberg2012}. Therefore, the population of subhalos orbiting a galaxy can be characterized using measurements of the surface density of stellar streams. Up to know, this analysis has been reduced to the Pal 5 and GD-1 stellar streams (see {\it e.g.} \cite{Carlberg2012, Bonaca:2018fek, Banik:2019cza}).

The number of known stellar streams has significantly increased in the past 5 years due to the advent of large sky photometric and astrometric surveys, such as the Sloan Digital Sky Survey, Dark Energy Survey, Pan-STARRS1 and Gaia. A compilation of known stellar streams can be found in the {\tt galstream} python package \cite{2018MNRAS.474.4112M}\footnote{\href{https://github.com/cmateu/galstreams}{https://github.com/cmateu/galstreams}.}.
However, only for roughly 10\% of these streams - for those that have been spectroscopically followed up - properties about their orbits and progenitors can be inferred. The near-future spectroscopic facilities (such as, 4MOST or DESI), are expected to increase, on the one hand, the number of cold tidal streams with known orbit and kinematics and, on the other hand, the number of member stars of individual streams, thus reducing measurement errors on the surface density. These improvements might lead to a sensitivity to the effects of subhalos with masses as low as $10^5\,\rm M_{\odot}$ \cite{Yoon+2011, Carlberg2012, Bovy:2016irg}.  To fully take advantage of the potential of stellar stream observations,  experimental advances need to be accompanied by improvements on the theoretical modeling of streams \cite{2020arXiv200107215B, Morinaga:2019fsy}.

The most stringent bound that could be imposed in this case is thus $\Mhm \lesssim 10^5 \Msun$ and therefore WDM, SIDM and FDM scenarios with parameter values up to
\bea
	m_\WDM &\sim \SI{40}{keV}, \\
	 \Tkd &\sim \SI{9}{keV}, \\
	m_\FDM &\sim \SI{2e-18}{eV},
\eea
may be probed.  Since FDM models could alleviate small-scale discrepancies when $m_{\rm FDM} \lesssim  \SI{e-21}{eV}$, realizing the full potential of from stellar stream observations can certainly test the FDM explanation of the small-scale problems. A similar conclusion can be drawn for WDM and SIDM models, especially when the resolution of the small scale discrepancies rests on the early suppression of the linear power spectrum.
 
\section{Discussion and conclusions}
\label{sec:Conclusions}

In this paper we considered different DM scenarios in the light of the recent indirectly measurement of dark subhalo abundance from the stellar density fluctuations of tidal streams. Stellar streams provide a promising set of observables for constraining small-scale structure and thus, studying the DM particle nature owing to the differences in subhalo abundances predicted by different DM models.  As shown in section~\ref{sec:streams}, current stellar stream limits are already at the level of the ones obtained from other observables, such as Lyman-$\alpha$ or gravitational lensing. In the conservative case, for WDM we obtain the lower bound $m_\WDM > \SI{3.2}{keV}$ which is slightly lower that found~\cite{Banik:2019cza}. Similarly, we find that the FDM mass must satisfy $m_\FDM > \SI{5.2e-21}{eV}$ and for SIDM scenarios where DM is coupled to a dark radiation bath, kinetic decoupling must take place at temperatures higher than $\Tkd > \SI{0.7}{keV}$.

A significant improvement on these bounds can be expected in the near future with the advent of large sky photometric and spectroscopic surveys which will increase the number of known streams and the number of member stars belonging to a given stellar stream.  In order to fully realize the potential of these measurements,  it would be necessary to improve our understanding of baryonic effects in disrupting DM subhalos. Also, an improved theoretical understanding of the substructure in SIDM models is required in order to effectively study them using stellar dynamics measurements.

Tidal stripping and disruption of subhalos due to baryonic effects is difficult to quantify. Modifications in the SHMF induced by these effects depend on the density profile of the subhalos, with cored profiles more prone to disruption~\cite{Errani&Penarrubia2019}. The latter itself is dependent on the underlying DM physics. In our simplified framework we parametrize our lack of knowledge through the survival fraction $f$ that can take values in a relatively wide range (see {\it e.g.}~\cite{Onguia+2009, Sawala+2017, GarrisonKimmel+2017}). The most conservative bounds for the modified DM scenarios can be obtained by choosing $f=1$. In order to distinguish different DM scenarios, on the other hand, an accurate estimate of the baryonic effects on the suppression of the subhalo abundance is required. Moreover, if the nature of DM would be determined from other observables, then stellar stream observations may help to precisely determine the baryonic effects {\it i.e.},  the value of $f$ in our simplified framework.

At any rate, since the different methods that aim to determine the number of subhalos are affected by different sources of systematic error, it is the complementarity of them that would allow to set robust constraints on the particle nature of DM.

The population of dark subhalos in our Galaxy further affects indirect DM searches which aim at detecting the flux of final stable particles produced by annihilation or decay of Weakly Interacting Massive Particles. The flux of stable particles depends on the distribution of DM in the region under study and the presence of subhalos can significantly boost this signal~\cite{2017MNRAS.466.4974M}. The subhalo abundance also impacts searches for annihilation signals from optically faint patches of the sky~\cite{2019Galax...7...90C}.

\emph{Note added:} During completion of this work a related study on FDM~\cite{Schutz:2020jox} appeared, where similar conclusions regarding FDM are reached using a slightly different modification of the SHMF (see also appendix~\ref{sec:AppendixA}).

\section*{Acknowledgements}
The authors thank Katelin Schutz, Mark Lovell, Matteo Viel and Nils Sch\"oneberg for helpful discussions. This work was supported by the European Regional Development Fund through the CoE program grant TK133, the Mobilitas Pluss grants MOBTP135, MOBTT5, MOBTT86, MOBJD323 and by the Estonian Research Council grant PRG803.

\appendix
\section{Analytic fitting forms for WDM \& FDM}
\label{sec:AppendixA}

In the following we assume analytic description for the SHMF as presented in~\cite{Schneider:2014rda}, where it has been extensively tested against WDM and mixedDM (warm+cold) N-body simulations with a wide range of effective small-scale power suppression scales. Even though a precise treatment for the FDM would require numerical integration of the Scr\"odinger-Poisson system in place of the usual Vlasov-Poisson system of the collisionless N-body problem, the above analytic model should still serve as a reasonably good approximation. The linear input spectra for the SHMF calculations were calculated with CAMB\footnote{\url{https://github.com/cmbant/CAMB}}\cite{Lewis:1999bs} and axionCAMB\footnote{\url{https://github.com/dgrin1/axionCAMB}}\cite{Hlozek:2014lca} Boltzmann codes (for WDM and FDM, respectively). Spatially flat cosmologies with $\Omega_m=0.3$, $\Omega_{\rm WDM/FDM}=0.25$, $h=0.7$ were assumed.

It turns out that the results of our analytical calculations can be quite well approximated by the following functional form
\bea\label{eq:fitform}
\frac{dn_{\rm sub}}{dM} & = R(M;\alpha,\beta,\gamma) \left(\frac{dn_{\rm sub}}{dM}\right)_{\text{CDM}}\\
R(M;\alpha,\beta,\gamma)&\equiv\left[1 + \left(\frac{\gamma\Mhm}{M}\right)^\alpha\right]^{-\beta}\,,
\eea
with best-fit parameters
\begin{itemize}
\item WDM:
  $\alpha\simeq 0.76$, $\beta\simeq 3.9$, $\gamma\simeq 0.087$,
\item FDM:
  $\alpha\simeq 1.5$, $\beta\simeq 5.0$, $\gamma\simeq 0.14$\,.
\end{itemize}
The above WDM fit is fine for $m_{\rm WDM}\sim 4-8$~keV but it can be significantly improved for lower and higher masses by allowing $\beta$ to vary with $m_{\rm WDM}$
\begin{eqnarray*}
  \beta &\simeq& 3.14 - 0.057\times m_{\rm WDM}[{\rm keV}]\,,\\
  \alpha &\simeq& 0.785\,,\\
  \gamma &\simeq& 0.153\,.
\end{eqnarray*}
Here the half-mode mass $\Mhm$ is given by eq.~\eqref{eq:mhm} where $\Mhm$ for WDM is shown in eq.~\eqref{eq:khm-wdm} while for FDM we use the following slightly improved fit over the one provided in~\cite{Hu:2000ke}:
\be
\label{ec:khm_updated}
\Mhm\simeq 6.3\times10^{10}\,{\rm M_{\odot}}\left(\frac{\Omega_m}{0.3}\right)\left(\frac{h}{0.7}\right)^2
\left(\frac{m_{\text{FDM}}}{\SI{e-22}{eV}}\right)^{-1.38}.
\ee

\begin{figure}
  \centering
  \includegraphics[width=0.5\textwidth]{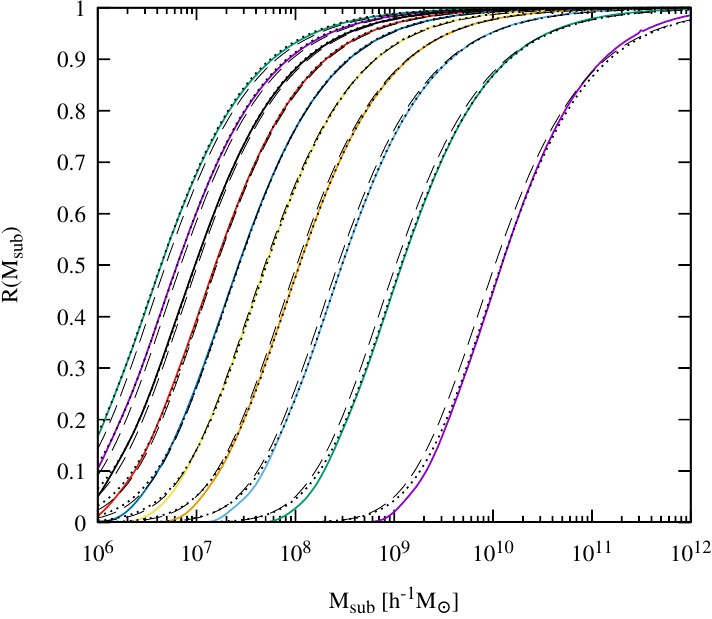}
  \caption{Ratio of WDM and CDM SHMFs. The WDM particle mass is in range $1-10$~keV with $1$~keV step size. Fitting functions \eqref{eq:fitform} with constant and mass-varying $\beta$ are shown with dashed and dotted lines, respectively.}\label{fig:WDM_fit}
\end{figure}

\begin{figure}
  \centering
  \includegraphics[width=0.5\textwidth]{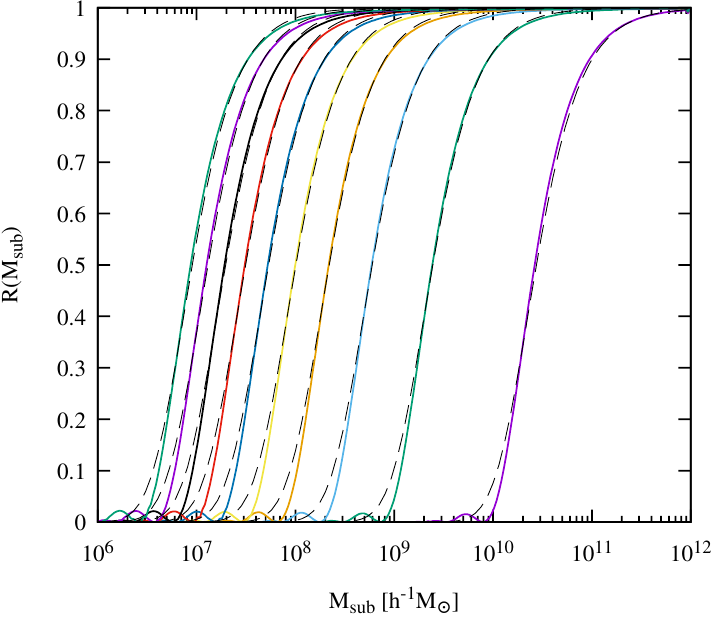}
  \caption{Ratio of FDM and CDM SHMFs. Here the FDM particle masses are obtained from the WDM masses of figure~\ref{fig:WDM_fit} via the approximate relation \eqref{eq:mwdm_2_mfdm} and cover a range $\sim (1-300)\times 10^{-22}$~eV. Fitting function \eqref{eq:fitform} is shown with dashed lines.}\label{fig:FDM_fit}
\end{figure}

The performance of the fitting form \eqref{eq:fitform} against direct analytic calculations is shown in figures~\ref{fig:WDM_fit} and~\ref{fig:FDM_fit} for WDM and FDM, respectively. Here the solid lines correspond to direct analytic calculations and dashed ones show the approximation form \eqref{eq:fitform}. In figure~\ref{fig:WDM_fit}, starting from the line with the strongest suppression, the WDM mass increases from $m_{\rm WDM}=1$~keV up to $10$~keV with $1$~keV step size. The FDM masses in figure~\ref{fig:FDM_fit} correspond to the WDM particle masses of figure~\ref{fig:WDM_fit} by requiring the linear power suppression wavenumbers $\khm$ to approximately match. In particular, this amounts to choosing
\be\label{eq:mwdm_2_mfdm}
m_{\rm FDM}\sim \left(\frac{m_{\rm WDM}}{{\rm keV}}\right)^{2.5}\times 10^{-22}\,{\rm eV}\,.
\ee

In a recent paper~\cite{Schutz:2020jox}, which studied topics similar to the ones covered in this work, the author used an analytic FDM substructure fitting function obtained earlier in~\cite{Du:2018wxl}. It turns out that our FDM results shown in figure~\ref{fig:FDM_fit} have somewhat stronger suppression. The results can be made to agree roughly if we increase $m_{\rm FDM}$ by a factor $\sim 4-5$. One possibility for this discrepancy could be the form of the filtering function used in calculations: Fourier \emph{vs} real-space top-hat function. In~\cite{Schneider:2014rda}, in context of WDM models, it was extensively discussed that standard Press-Schechter calculations with real-space top-hat function leads to a significant underestimation of the small-scale substructure suppression factor when compared with direct N-body results. However, it turns out that this is not the case here, since the fitting functions of~\cite{Du:2018wxl} also assumed sharp Fourier-space filter. Further studies are needed to resolve this discrepancy. For a possible explanation see the updated version of~\cite{Schutz:2020jox}. In general, one needs dedicated high resolution numerical simulations to reliably calibrate the analytic SHMF. Unfortunately, currently available simulations lack resolution to reach conclusive results.

\section{Comparison with \href{https://arxiv.org/abs/2003.01125}{arXiv:2003.01125}}
\label{sec:AppendixB}

During the reviewing process of this paper a simulation work~\cite{Lovell:2020bcy} appeared, which gave the results for the WDM SHMFs using identical fitting form to the one adopted here, {\emph cf.}, \eqref{eq:fitform}. It is instructive to see how our results compare. At first sight, the best-fit parameter values obtained there
\[
\alpha\simeq 2.5,\, \beta\simeq 0.2,\, \gamma\simeq 4.2
\]
seem to differ remarkably from ours:
\[
\alpha\simeq 0.76,\, \beta\simeq 3.9,\, \gamma\simeq 0.087h^{-1} \simeq 0.12\,.
\]
Note that we have adjusted our $\gamma$ parameter to count for the fact that in~\cite{Lovell:2020bcy} mass was measured in plain Solar masses, rather than in $h^{-1}\Msun$, as is assumed throughout this work. 

\begin{figure}
  \centering
  \includegraphics[width=0.5\textwidth]{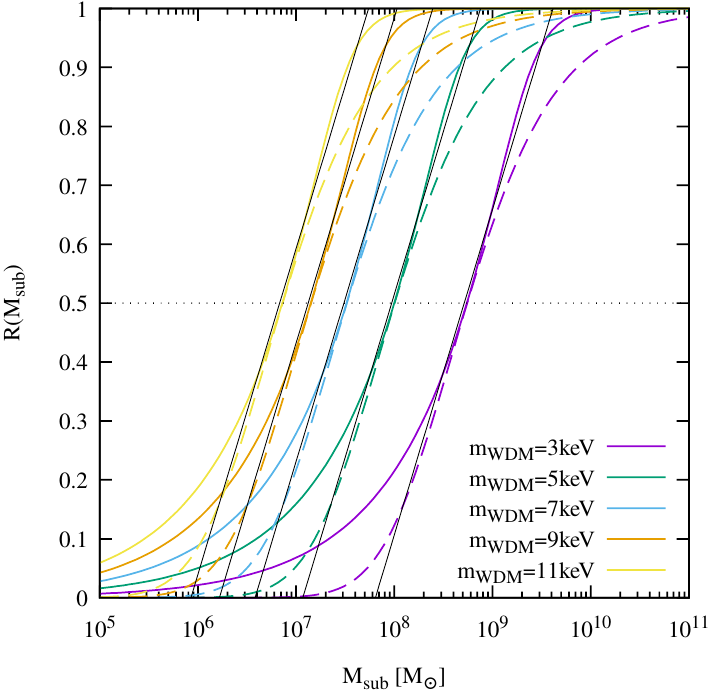}
  \caption{Comparison of substructure suppression factors. Solid and long-dashed lines present fits obtained in~\cite{Lovell:2020bcy} and in this work, respectively. $\Mhm$ corresponds to $m_\WDM$ values as given in the legend. Thin black lines display approximating function Eq.~(\ref{approxfunc}) (with $\Mhalf=\Mhm$).}
  \label{fig:lovell_compare}
\end{figure}

Here we want to show that for substructure suppression factors $R\sim 0.3-0.7$ both fits agree remarkably well. To see this let us re-express Eq.~(\ref{eq:fitform})
\be
R(M)=\left[1+(2^{1/\beta}-1)\left(\frac{\Mhalf}{M}\right)^\alpha\right]^{-\beta}\,,
\ee
where $\Mhalf$ is defined via $R(\Mhalf)\equiv \sfrac{1}{2}$.
It is related to $\Mhm$ as
\be
\Mhalf=\frac{\gamma}{(2^{1/\beta}-1)^{1/\alpha}}\Mhm\,.
\ee
Around masses $\Mhalf$ the substructure suppression factor $R(M)$ is well approximated by a linear relation in $R(M)$-$\ln(M)$ axes
\bea\label{approxfunc}
R(M)&\simeq \frac{1}{2} + n\cdot\ln\left(\frac{M}{\Mhalf}\right)\,,\\
n&\equiv \frac{1}{2}\alpha\beta\left(1-2^{-1/\beta}\right)\,.
\eea
It is easy to check that both sets of $\alpha$, $\beta$ and $\gamma$ parameters, as presented above, give practically the same values for $n$ and $\Mhalf$:
\begin{itemize}
\item our best-fit parameters result in: \\$\Mhalf\simeq 1.03\Mhm$, $n\simeq 0.24$\,,
\item whereas parameters of~\cite{Lovell:2020bcy} give:\\ $\Mhalf\simeq 1.06\Mhm$, $n\simeq 0.24$\,.
\end{itemize}
Thus, around $\Mhalf$ both fits are practically identical. This is illustrated in Figure~\ref{fig:lovell_compare}.

It is quite remarkable that the WDM fitting formulas of the current work, which are based on the analytic formalism of~\cite{Schneider:2014rda}, agree so well with the N-body results of~\cite{Lovell:2020bcy} down to substructure suppression factors of $R\sim 1/2$ and below. One also has to bear in mind that the values for the best-fit $\alpha$, $\beta$ and $\gamma$ parameters obtained in this work were largely driven by the behaviour of the function $R(M)$ at large values of $M$, which are hard to probe in limited set of N-body simulations. Additionally, at small masses simulations have lots of small-scale clumps out of which a significant fraction might be spurious systems.

\bibliographystyle{apsrev4-1}
\bibliography{dark_subs}

\begin{thebibliography}{112}%
\makeatletter
\providecommand \@ifxundefined [1]{%
 \@ifx{#1\undefined}
}%
\providecommand \@ifnum [1]{%
 \ifnum #1\expandafter \@firstoftwo
 \else \expandafter \@secondoftwo
 \fi
}%
\providecommand \@ifx [1]{%
 \ifx #1\expandafter \@firstoftwo
 \else \expandafter \@secondoftwo
 \fi
}%
\providecommand \natexlab [1]{#1}%
\providecommand \enquote  [1]{``#1''}%
\providecommand \bibnamefont  [1]{#1}%
\providecommand \bibfnamefont [1]{#1}%
\providecommand \citenamefont [1]{#1}%
\providecommand \href@noop [0]{\@secondoftwo}%
\providecommand \href [0]{\begingroup \@sanitize@url \@href}%
\providecommand \@href[1]{\@@startlink{#1}\@@href}%
\providecommand \@@href[1]{\endgroup#1\@@endlink}%
\providecommand \@sanitize@url [0]{\catcode `\\12\catcode `\$12\catcode
  `\&12\catcode `\#12\catcode `\^12\catcode `\_12\catcode `\%12\relax}%
\providecommand \@@startlink[1]{}%
\providecommand \@@endlink[0]{}%
\providecommand \url  [0]{\begingroup\@sanitize@url \@url }%
\providecommand \@url [1]{\endgroup\@href {#1}{\urlprefix }}%
\providecommand \urlprefix  [0]{URL }%
\providecommand \Eprint [0]{\href }%
\providecommand \doibase [0]{http://dx.doi.org/}%
\providecommand \selectlanguage [0]{\@gobble}%
\providecommand \bibinfo  [0]{\@secondoftwo}%
\providecommand \bibfield  [0]{\@secondoftwo}%
\providecommand \translation [1]{[#1]}%
\providecommand \BibitemOpen [0]{}%
\providecommand \bibitemStop [0]{}%
\providecommand \bibitemNoStop [0]{.\EOS\space}%
\providecommand \EOS [0]{\spacefactor3000\relax}%
\providecommand \BibitemShut  [1]{\csname bibitem#1\endcsname}%
\let\auto@bib@innerbib\@empty
\bibitem [{\citenamefont {Jungman}\ \emph {et~al.}(1996)\citenamefont
  {Jungman}, \citenamefont {Kamionkowski},\ and\ \citenamefont
  {Griest}}]{Jungman:1995df}%
  \BibitemOpen
  \bibfield  {author} {\bibinfo {author} {\bibfnamefont {G.}~\bibnamefont
  {Jungman}}, \bibinfo {author} {\bibfnamefont {M.}~\bibnamefont
  {Kamionkowski}}, \ and\ \bibinfo {author} {\bibfnamefont {K.}~\bibnamefont
  {Griest}},\ }\href {\doibase 10.1016/0370-1573(95)00058-5} {\bibfield
  {journal} {\bibinfo  {journal} {Phys.Rept.}\ }\textbf {\bibinfo {volume}
  {267}},\ \bibinfo {pages} {195} (\bibinfo {year} {1996})},\ \Eprint
  {http://arxiv.org/abs/hep-ph/9506380} {arXiv:hep-ph/9506380} \BibitemShut
  {NoStop}%
\bibitem [{\citenamefont {Bertone}\ \emph {et~al.}(2005)\citenamefont
  {Bertone}, \citenamefont {Hooper},\ and\ \citenamefont
  {Silk}}]{Bertone:2004pz}%
  \BibitemOpen
  \bibfield  {author} {\bibinfo {author} {\bibfnamefont {G.}~\bibnamefont
  {Bertone}}, \bibinfo {author} {\bibfnamefont {D.}~\bibnamefont {Hooper}}, \
  and\ \bibinfo {author} {\bibfnamefont {J.}~\bibnamefont {Silk}},\ }\href
  {\doibase 10.1016/j.physrep.2004.08.031} {\bibfield  {journal} {\bibinfo
  {journal} {Phys.Rept.}\ }\textbf {\bibinfo {volume} {405}},\ \bibinfo {pages}
  {279} (\bibinfo {year} {2005})},\ \Eprint
  {http://arxiv.org/abs/hep-ph/0404175} {arXiv:hep-ph/0404175} \BibitemShut
  {NoStop}%
\bibitem [{\citenamefont {Klypin}\ \emph {et~al.}(1999)\citenamefont {Klypin},
  \citenamefont {Kravtsov}, \citenamefont {Valenzuela},\ and\ \citenamefont
  {Prada}}]{Klypin:1999uc}%
  \BibitemOpen
  \bibfield  {author} {\bibinfo {author} {\bibfnamefont {A.~A.}\ \bibnamefont
  {Klypin}}, \bibinfo {author} {\bibfnamefont {A.~V.}\ \bibnamefont
  {Kravtsov}}, \bibinfo {author} {\bibfnamefont {O.}~\bibnamefont
  {Valenzuela}}, \ and\ \bibinfo {author} {\bibfnamefont {F.}~\bibnamefont
  {Prada}},\ }\href {\doibase 10.1086/307643} {\bibfield  {journal} {\bibinfo
  {journal} {Astrophys. J.}\ }\textbf {\bibinfo {volume} {522}},\ \bibinfo
  {pages} {82} (\bibinfo {year} {1999})},\ \Eprint
  {http://arxiv.org/abs/astro-ph/9901240} {arXiv:astro-ph/9901240 [astro-ph]}
  \BibitemShut {NoStop}%
\bibitem [{\citenamefont {Moore}\ \emph
  {et~al.}(1999{\natexlab{a}})\citenamefont {Moore}, \citenamefont {Ghigna},
  \citenamefont {Governato}, \citenamefont {Lake}, \citenamefont {Quinn},
  \citenamefont {Stadel},\ and\ \citenamefont {Tozzi}}]{Moore:1999nt}%
  \BibitemOpen
  \bibfield  {author} {\bibinfo {author} {\bibfnamefont {B.}~\bibnamefont
  {Moore}}, \bibinfo {author} {\bibfnamefont {S.}~\bibnamefont {Ghigna}},
  \bibinfo {author} {\bibfnamefont {F.}~\bibnamefont {Governato}}, \bibinfo
  {author} {\bibfnamefont {G.}~\bibnamefont {Lake}}, \bibinfo {author}
  {\bibfnamefont {T.~R.}\ \bibnamefont {Quinn}}, \bibinfo {author}
  {\bibfnamefont {J.}~\bibnamefont {Stadel}}, \ and\ \bibinfo {author}
  {\bibfnamefont {P.}~\bibnamefont {Tozzi}},\ }\href {\doibase 10.1086/312287}
  {\bibfield  {journal} {\bibinfo  {journal} {Astrophys. J.}\ }\textbf
  {\bibinfo {volume} {524}},\ \bibinfo {pages} {L19} (\bibinfo {year}
  {1999}{\natexlab{a}})},\ \Eprint {http://arxiv.org/abs/astro-ph/9907411}
  {arXiv:astro-ph/9907411 [astro-ph]} \BibitemShut {NoStop}%
\bibitem [{\citenamefont {Boylan-Kolchin}\ \emph {et~al.}(2011)\citenamefont
  {Boylan-Kolchin}, \citenamefont {Bullock},\ and\ \citenamefont
  {Kaplinghat}}]{BoylanKolchin:2011de}%
  \BibitemOpen
  \bibfield  {author} {\bibinfo {author} {\bibfnamefont {M.}~\bibnamefont
  {Boylan-Kolchin}}, \bibinfo {author} {\bibfnamefont {J.~S.}\ \bibnamefont
  {Bullock}}, \ and\ \bibinfo {author} {\bibfnamefont {M.}~\bibnamefont
  {Kaplinghat}},\ }\href {\doibase 10.1111/j.1745-3933.2011.01074.x} {\bibfield
   {journal} {\bibinfo  {journal} {Mon. Not. Roy. Astron. Soc.}\ }\textbf
  {\bibinfo {volume} {415}},\ \bibinfo {pages} {L40} (\bibinfo {year}
  {2011})},\ \Eprint {http://arxiv.org/abs/1103.0007} {arXiv:1103.0007
  [astro-ph.CO]} \BibitemShut {NoStop}%
\bibitem [{\citenamefont {Boylan-Kolchin}\ \emph {et~al.}(2012)\citenamefont
  {Boylan-Kolchin}, \citenamefont {Bullock},\ and\ \citenamefont
  {Kaplinghat}}]{BoylanKolchin:2011dk}%
  \BibitemOpen
  \bibfield  {author} {\bibinfo {author} {\bibfnamefont {M.}~\bibnamefont
  {Boylan-Kolchin}}, \bibinfo {author} {\bibfnamefont {J.~S.}\ \bibnamefont
  {Bullock}}, \ and\ \bibinfo {author} {\bibfnamefont {M.}~\bibnamefont
  {Kaplinghat}},\ }\href {\doibase 10.1111/j.1365-2966.2012.20695.x} {\bibfield
   {journal} {\bibinfo  {journal} {Mon. Not. Roy. Astron. Soc.}\ }\textbf
  {\bibinfo {volume} {422}},\ \bibinfo {pages} {1203} (\bibinfo {year}
  {2012})},\ \Eprint {http://arxiv.org/abs/1111.2048} {arXiv:1111.2048
  [astro-ph.CO]} \BibitemShut {NoStop}%
\bibitem [{\citenamefont {Garrison-Kimmel}\ \emph {et~al.}(2014)\citenamefont
  {Garrison-Kimmel}, \citenamefont {Boylan-Kolchin}, \citenamefont {Bullock},\
  and\ \citenamefont {Kirby}}]{Garrison-Kimmel:2014vqa}%
  \BibitemOpen
  \bibfield  {author} {\bibinfo {author} {\bibfnamefont {S.}~\bibnamefont
  {Garrison-Kimmel}}, \bibinfo {author} {\bibfnamefont {M.}~\bibnamefont
  {Boylan-Kolchin}}, \bibinfo {author} {\bibfnamefont {J.~S.}\ \bibnamefont
  {Bullock}}, \ and\ \bibinfo {author} {\bibfnamefont {E.~N.}\ \bibnamefont
  {Kirby}},\ }\href {\doibase 10.1093/mnras/stu1477} {\bibfield  {journal}
  {\bibinfo  {journal} {Mon. Not. Roy. Astron. Soc.}\ }\textbf {\bibinfo
  {volume} {444}},\ \bibinfo {pages} {222} (\bibinfo {year} {2014})},\ \Eprint
  {http://arxiv.org/abs/1404.5313} {arXiv:1404.5313 [astro-ph.GA]} \BibitemShut
  {NoStop}%
\bibitem [{\citenamefont {Papastergis}\ \emph {et~al.}(2015)\citenamefont
  {Papastergis}, \citenamefont {Giovanelli}, \citenamefont {Haynes},\ and\
  \citenamefont {Shankar}}]{Papastergis:2014aba}%
  \BibitemOpen
  \bibfield  {author} {\bibinfo {author} {\bibfnamefont {E.}~\bibnamefont
  {Papastergis}}, \bibinfo {author} {\bibfnamefont {R.}~\bibnamefont
  {Giovanelli}}, \bibinfo {author} {\bibfnamefont {M.~P.}\ \bibnamefont
  {Haynes}}, \ and\ \bibinfo {author} {\bibfnamefont {F.}~\bibnamefont
  {Shankar}},\ }\href {\doibase 10.1051/0004-6361/201424909} {\bibfield
  {journal} {\bibinfo  {journal} {Astron. Astrophys.}\ }\textbf {\bibinfo
  {volume} {574}},\ \bibinfo {pages} {A113} (\bibinfo {year} {2015})},\ \Eprint
  {http://arxiv.org/abs/1407.4665} {arXiv:1407.4665 [astro-ph.GA]} \BibitemShut
  {NoStop}%
\bibitem [{\citenamefont {Kaplinghat}\ \emph {et~al.}(2019)\citenamefont
  {Kaplinghat}, \citenamefont {Valli},\ and\ \citenamefont
  {Yu}}]{Kaplinghat:2019svz}%
  \BibitemOpen
  \bibfield  {author} {\bibinfo {author} {\bibfnamefont {M.}~\bibnamefont
  {Kaplinghat}}, \bibinfo {author} {\bibfnamefont {M.}~\bibnamefont {Valli}}, \
  and\ \bibinfo {author} {\bibfnamefont {H.-B.}\ \bibnamefont {Yu}},\ }\href
  {\doibase 10.1093/mnras/stz2511} {\bibfield  {journal} {\bibinfo  {journal}
  {Mon. Not. Roy. Astron. Soc.}\ }\textbf {\bibinfo {volume} {490}},\ \bibinfo
  {pages} {231} (\bibinfo {year} {2019})},\ \Eprint
  {http://arxiv.org/abs/1904.04939} {arXiv:1904.04939 [astro-ph.GA]}
  \BibitemShut {NoStop}%
\bibitem [{\citenamefont {Moore}(1994)}]{Moore:1994yx}%
  \BibitemOpen
  \bibfield  {author} {\bibinfo {author} {\bibfnamefont {B.}~\bibnamefont
  {Moore}},\ }\href {\doibase 10.1038/370629a0} {\bibfield  {journal} {\bibinfo
   {journal} {Nature}\ }\textbf {\bibinfo {volume} {370}},\ \bibinfo {pages}
  {629} (\bibinfo {year} {1994})}\BibitemShut {NoStop}%
\bibitem [{\citenamefont {Flores}\ and\ \citenamefont
  {Primack}(1994)}]{Flores:1994gz}%
  \BibitemOpen
  \bibfield  {author} {\bibinfo {author} {\bibfnamefont {R.~A.}\ \bibnamefont
  {Flores}}\ and\ \bibinfo {author} {\bibfnamefont {J.~R.}\ \bibnamefont
  {Primack}},\ }\href {\doibase 10.1086/187350} {\bibfield  {journal} {\bibinfo
   {journal} {Astrophys. J.}\ }\textbf {\bibinfo {volume} {427}},\ \bibinfo
  {pages} {L1} (\bibinfo {year} {1994})},\ \Eprint
  {http://arxiv.org/abs/astro-ph/9402004} {arXiv:astro-ph/9402004 [astro-ph]}
  \BibitemShut {NoStop}%
\bibitem [{\citenamefont {Burkert}(1996)}]{Burkert:1995yz}%
  \BibitemOpen
  \bibfield  {author} {\bibinfo {author} {\bibfnamefont {A.}~\bibnamefont
  {Burkert}},\ }\bibfield  {booktitle} {\emph {\bibinfo {booktitle} {{IAU
  Symposium 171: New Light on Galaxy Evolution Heidelberg, Germany, June 26-30,
  1995}}},\ }\href {\doibase 10.1086/309560} {\bibfield  {journal} {\bibinfo
  {journal} {IAU Symp.}\ }\textbf {\bibinfo {volume} {171}},\ \bibinfo {pages}
  {175} (\bibinfo {year} {1996})},\ \bibinfo {note} {[Astrophys.
  J.447,L25(1995)]},\ \Eprint {http://arxiv.org/abs/astro-ph/9504041}
  {arXiv:astro-ph/9504041 [astro-ph]} \BibitemShut {NoStop}%
\bibitem [{\citenamefont {Moore}\ \emph
  {et~al.}(1999{\natexlab{b}})\citenamefont {Moore}, \citenamefont {Quinn},
  \citenamefont {Governato}, \citenamefont {Stadel},\ and\ \citenamefont
  {Lake}}]{Moore:1999gc}%
  \BibitemOpen
  \bibfield  {author} {\bibinfo {author} {\bibfnamefont {B.}~\bibnamefont
  {Moore}}, \bibinfo {author} {\bibfnamefont {T.~R.}\ \bibnamefont {Quinn}},
  \bibinfo {author} {\bibfnamefont {F.}~\bibnamefont {Governato}}, \bibinfo
  {author} {\bibfnamefont {J.}~\bibnamefont {Stadel}}, \ and\ \bibinfo {author}
  {\bibfnamefont {G.}~\bibnamefont {Lake}},\ }\href {\doibase
  10.1046/j.1365-8711.1999.03039.x} {\bibfield  {journal} {\bibinfo  {journal}
  {Mon. Not. Roy. Astron. Soc.}\ }\textbf {\bibinfo {volume} {310}},\ \bibinfo
  {pages} {1147} (\bibinfo {year} {1999}{\natexlab{b}})},\ \Eprint
  {http://arxiv.org/abs/astro-ph/9903164} {arXiv:astro-ph/9903164 [astro-ph]}
  \BibitemShut {NoStop}%
\bibitem [{\citenamefont {van~den Bosch}\ and\ \citenamefont
  {Swaters}(2001)}]{vandenBosch:2000rza}%
  \BibitemOpen
  \bibfield  {author} {\bibinfo {author} {\bibfnamefont {F.~C.}\ \bibnamefont
  {van~den Bosch}}\ and\ \bibinfo {author} {\bibfnamefont {R.~A.}\ \bibnamefont
  {Swaters}},\ }\href {\doibase 10.1046/j.1365-8711.2001.04456.x} {\bibfield
  {journal} {\bibinfo  {journal} {Mon. Not. Roy. Astron. Soc.}\ }\textbf
  {\bibinfo {volume} {325}},\ \bibinfo {pages} {1017} (\bibinfo {year}
  {2001})},\ \Eprint {http://arxiv.org/abs/astro-ph/0006048}
  {arXiv:astro-ph/0006048 [astro-ph]} \BibitemShut {NoStop}%
\bibitem [{\citenamefont {de~Blok}\ \emph {et~al.}(2001)\citenamefont
  {de~Blok}, \citenamefont {McGaugh}, \citenamefont {Bosma},\ and\
  \citenamefont {Rubin}}]{deBlok:2001hbg}%
  \BibitemOpen
  \bibfield  {author} {\bibinfo {author} {\bibfnamefont {W.~J.~G.}\
  \bibnamefont {de~Blok}}, \bibinfo {author} {\bibfnamefont {S.~S.}\
  \bibnamefont {McGaugh}}, \bibinfo {author} {\bibfnamefont {A.}~\bibnamefont
  {Bosma}}, \ and\ \bibinfo {author} {\bibfnamefont {V.~C.}\ \bibnamefont
  {Rubin}},\ }\href {\doibase 10.1086/320262} {\bibfield  {journal} {\bibinfo
  {journal} {Astrophys. J.}\ }\textbf {\bibinfo {volume} {552}},\ \bibinfo
  {pages} {L23} (\bibinfo {year} {2001})},\ \Eprint
  {http://arxiv.org/abs/astro-ph/0103102} {arXiv:astro-ph/0103102 [astro-ph]}
  \BibitemShut {NoStop}%
\bibitem [{\citenamefont {Pawlowski}\ \emph {et~al.}(2013)\citenamefont
  {Pawlowski}, \citenamefont {Kroupa},\ and\ \citenamefont
  {Jerjen}}]{Pawlowski:2013kpa}%
  \BibitemOpen
  \bibfield  {author} {\bibinfo {author} {\bibfnamefont {M.~S.}\ \bibnamefont
  {Pawlowski}}, \bibinfo {author} {\bibfnamefont {P.}~\bibnamefont {Kroupa}}, \
  and\ \bibinfo {author} {\bibfnamefont {H.}~\bibnamefont {Jerjen}},\ }\href
  {\doibase 10.1093/mnras/stt1384} {\bibfield  {journal} {\bibinfo  {journal}
  {Mon. Not. Roy. Astron. Soc.}\ }\textbf {\bibinfo {volume} {435}},\ \bibinfo
  {pages} {1928} (\bibinfo {year} {2013})},\ \Eprint
  {http://arxiv.org/abs/1307.6210} {arXiv:1307.6210 [astro-ph.CO]} \BibitemShut
  {NoStop}%
\bibitem [{\citenamefont {Oman}\ \emph {et~al.}(2015)\citenamefont {Oman} \emph
  {et~al.}}]{Oman:2015xda}%
  \BibitemOpen
  \bibfield  {author} {\bibinfo {author} {\bibfnamefont {K.~A.}\ \bibnamefont
  {Oman}} \emph {et~al.},\ }\href {\doibase 10.1093/mnras/stv1504} {\bibfield
  {journal} {\bibinfo  {journal} {Mon. Not. Roy. Astron. Soc.}\ }\textbf
  {\bibinfo {volume} {452}},\ \bibinfo {pages} {3650} (\bibinfo {year}
  {2015})},\ \Eprint {http://arxiv.org/abs/1504.01437} {arXiv:1504.01437
  [astro-ph.GA]} \BibitemShut {NoStop}%
\bibitem [{\citenamefont {Kuhlen}\ \emph {et~al.}(2012)\citenamefont {Kuhlen},
  \citenamefont {Vogelsberger},\ and\ \citenamefont {Angulo}}]{Kuhlen:2012ft}%
  \BibitemOpen
  \bibfield  {author} {\bibinfo {author} {\bibfnamefont {M.}~\bibnamefont
  {Kuhlen}}, \bibinfo {author} {\bibfnamefont {M.}~\bibnamefont
  {Vogelsberger}}, \ and\ \bibinfo {author} {\bibfnamefont {R.}~\bibnamefont
  {Angulo}},\ }\href {\doibase 10.1016/j.dark.2012.10.002} {\bibfield
  {journal} {\bibinfo  {journal} {Phys. Dark Univ.}\ }\textbf {\bibinfo
  {volume} {1}},\ \bibinfo {pages} {50} (\bibinfo {year} {2012})},\ \Eprint
  {http://arxiv.org/abs/1209.5745} {arXiv:1209.5745 [astro-ph.CO]} \BibitemShut
  {NoStop}%
\bibitem [{\citenamefont {Tulin}\ and\ \citenamefont
  {Yu}(2018)}]{Tulin:2017ara}%
  \BibitemOpen
  \bibfield  {author} {\bibinfo {author} {\bibfnamefont {S.}~\bibnamefont
  {Tulin}}\ and\ \bibinfo {author} {\bibfnamefont {H.-B.}\ \bibnamefont {Yu}},\
  }\href {\doibase 10.1016/j.physrep.2017.11.004} {\bibfield  {journal}
  {\bibinfo  {journal} {Phys. Rept.}\ }\textbf {\bibinfo {volume} {730}},\
  \bibinfo {pages} {1} (\bibinfo {year} {2018})},\ \Eprint
  {http://arxiv.org/abs/1705.02358} {arXiv:1705.02358 [hep-ph]} \BibitemShut
  {NoStop}%
\bibitem [{\citenamefont {{Bullock}}\ and\ \citenamefont
  {{Boylan-Kolchin}}(2017)}]{2017ARA&A..55..343B}%
  \BibitemOpen
  \bibfield  {author} {\bibinfo {author} {\bibfnamefont {J.~S.}\ \bibnamefont
  {{Bullock}}}\ and\ \bibinfo {author} {\bibfnamefont {M.}~\bibnamefont
  {{Boylan-Kolchin}}},\ }\href {\doibase 10.1146/annurev-astro-091916-055313}
  {\bibfield  {journal} {\bibinfo  {journal} {\araa}\ }\textbf {\bibinfo
  {volume} {55}},\ \bibinfo {pages} {343} (\bibinfo {year} {2017})},\ \Eprint
  {http://arxiv.org/abs/1707.04256} {arXiv:1707.04256 [astro-ph.CO]}
  \BibitemShut {NoStop}%
\bibitem [{\citenamefont {Bode}\ \emph {et~al.}(2001)\citenamefont {Bode},
  \citenamefont {Ostriker},\ and\ \citenamefont {Turok}}]{Bode:2000gq}%
  \BibitemOpen
  \bibfield  {author} {\bibinfo {author} {\bibfnamefont {P.}~\bibnamefont
  {Bode}}, \bibinfo {author} {\bibfnamefont {J.~P.}\ \bibnamefont {Ostriker}},
  \ and\ \bibinfo {author} {\bibfnamefont {N.}~\bibnamefont {Turok}},\ }\href
  {\doibase 10.1086/321541} {\bibfield  {journal} {\bibinfo  {journal}
  {Astrophys.J.}\ }\textbf {\bibinfo {volume} {556}},\ \bibinfo {pages} {93}
  (\bibinfo {year} {2001})},\ \Eprint {http://arxiv.org/abs/astro-ph/0010389}
  {arXiv:astro-ph/0010389} \BibitemShut {NoStop}%
\bibitem [{\citenamefont {Colin}\ \emph {et~al.}(2000)\citenamefont {Colin},
  \citenamefont {Avila-Reese},\ and\ \citenamefont
  {Valenzuela}}]{Colin:2000dn}%
  \BibitemOpen
  \bibfield  {author} {\bibinfo {author} {\bibfnamefont {P.}~\bibnamefont
  {Colin}}, \bibinfo {author} {\bibfnamefont {V.}~\bibnamefont {Avila-Reese}},
  \ and\ \bibinfo {author} {\bibfnamefont {O.}~\bibnamefont {Valenzuela}},\
  }\href {\doibase 10.1086/317057} {\bibfield  {journal} {\bibinfo  {journal}
  {Astrophys. J.}\ }\textbf {\bibinfo {volume} {542}},\ \bibinfo {pages} {622}
  (\bibinfo {year} {2000})},\ \Eprint {http://arxiv.org/abs/astro-ph/0004115}
  {arXiv:astro-ph/0004115 [astro-ph]} \BibitemShut {NoStop}%
\bibitem [{\citenamefont {Hu}\ \emph {et~al.}(2000)\citenamefont {Hu},
  \citenamefont {Barkana},\ and\ \citenamefont {Gruzinov}}]{Hu:2000ke}%
  \BibitemOpen
  \bibfield  {author} {\bibinfo {author} {\bibfnamefont {W.}~\bibnamefont
  {Hu}}, \bibinfo {author} {\bibfnamefont {R.}~\bibnamefont {Barkana}}, \ and\
  \bibinfo {author} {\bibfnamefont {A.}~\bibnamefont {Gruzinov}},\ }\href
  {\doibase 10.1103/PhysRevLett.85.1158} {\bibfield  {journal} {\bibinfo
  {journal} {Phys. Rev. Lett.}\ }\textbf {\bibinfo {volume} {85}},\ \bibinfo
  {pages} {1158} (\bibinfo {year} {2000})},\ \Eprint
  {http://arxiv.org/abs/astro-ph/0003365} {arXiv:astro-ph/0003365 [astro-ph]}
  \BibitemShut {NoStop}%
\bibitem [{\citenamefont {Schive}\ \emph {et~al.}(2016)\citenamefont {Schive},
  \citenamefont {Chiueh}, \citenamefont {Broadhurst},\ and\ \citenamefont
  {Huang}}]{Schive:2015kza}%
  \BibitemOpen
  \bibfield  {author} {\bibinfo {author} {\bibfnamefont {H.-Y.}\ \bibnamefont
  {Schive}}, \bibinfo {author} {\bibfnamefont {T.}~\bibnamefont {Chiueh}},
  \bibinfo {author} {\bibfnamefont {T.}~\bibnamefont {Broadhurst}}, \ and\
  \bibinfo {author} {\bibfnamefont {K.-W.}\ \bibnamefont {Huang}},\ }\href
  {\doibase 10.3847/0004-637X/818/1/89} {\bibfield  {journal} {\bibinfo
  {journal} {Astrophys. J.}\ }\textbf {\bibinfo {volume} {818}},\ \bibinfo
  {pages} {89} (\bibinfo {year} {2016})},\ \Eprint
  {http://arxiv.org/abs/1508.04621} {arXiv:1508.04621 [astro-ph.GA]}
  \BibitemShut {NoStop}%
\bibitem [{\citenamefont {Marsh}(2016{\natexlab{a}})}]{Marsh:2015xka}%
  \BibitemOpen
  \bibfield  {author} {\bibinfo {author} {\bibfnamefont {D.~J.~E.}\
  \bibnamefont {Marsh}},\ }\href {\doibase 10.1016/j.physrep.2016.06.005}
  {\bibfield  {journal} {\bibinfo  {journal} {Phys. Rept.}\ }\textbf {\bibinfo
  {volume} {643}},\ \bibinfo {pages} {1} (\bibinfo {year}
  {2016}{\natexlab{a}})},\ \Eprint {http://arxiv.org/abs/1510.07633}
  {arXiv:1510.07633 [astro-ph.CO]} \BibitemShut {NoStop}%
\bibitem [{\citenamefont {Hui}\ \emph {et~al.}(2017)\citenamefont {Hui},
  \citenamefont {Ostriker}, \citenamefont {Tremaine},\ and\ \citenamefont
  {Witten}}]{Hui:2016ltb}%
  \BibitemOpen
  \bibfield  {author} {\bibinfo {author} {\bibfnamefont {L.}~\bibnamefont
  {Hui}}, \bibinfo {author} {\bibfnamefont {J.~P.}\ \bibnamefont {Ostriker}},
  \bibinfo {author} {\bibfnamefont {S.}~\bibnamefont {Tremaine}}, \ and\
  \bibinfo {author} {\bibfnamefont {E.}~\bibnamefont {Witten}},\ }\href
  {\doibase 10.1103/PhysRevD.95.043541} {\bibfield  {journal} {\bibinfo
  {journal} {Phys. Rev.}\ }\textbf {\bibinfo {volume} {D95}},\ \bibinfo {pages}
  {043541} (\bibinfo {year} {2017})},\ \Eprint
  {http://arxiv.org/abs/1610.08297} {arXiv:1610.08297 [astro-ph.CO]}
  \BibitemShut {NoStop}%
\bibitem [{\citenamefont {Du}\ \emph {et~al.}(2017)\citenamefont {Du},
  \citenamefont {Behrens},\ and\ \citenamefont {Niemeyer}}]{Du:2016zcv}%
  \BibitemOpen
  \bibfield  {author} {\bibinfo {author} {\bibfnamefont {X.}~\bibnamefont
  {Du}}, \bibinfo {author} {\bibfnamefont {C.}~\bibnamefont {Behrens}}, \ and\
  \bibinfo {author} {\bibfnamefont {J.~C.}\ \bibnamefont {Niemeyer}},\ }\href
  {\doibase 10.1093/mnras/stw2724} {\bibfield  {journal} {\bibinfo  {journal}
  {Mon. Not. Roy. Astron. Soc.}\ }\textbf {\bibinfo {volume} {465}},\ \bibinfo
  {pages} {941} (\bibinfo {year} {2017})},\ \Eprint
  {http://arxiv.org/abs/1608.02575} {arXiv:1608.02575 [astro-ph.CO]}
  \BibitemShut {NoStop}%
\bibitem [{\citenamefont {Cyr-Racine}\ and\ \citenamefont
  {Sigurdson}(2013)}]{CyrRacine:2012fz}%
  \BibitemOpen
  \bibfield  {author} {\bibinfo {author} {\bibfnamefont {F.-Y.}\ \bibnamefont
  {Cyr-Racine}}\ and\ \bibinfo {author} {\bibfnamefont {K.}~\bibnamefont
  {Sigurdson}},\ }\href {\doibase 10.1103/PhysRevD.87.103515} {\bibfield
  {journal} {\bibinfo  {journal} {Phys. Rev.}\ }\textbf {\bibinfo {volume}
  {D87}},\ \bibinfo {pages} {103515} (\bibinfo {year} {2013})},\ \Eprint
  {http://arxiv.org/abs/1209.5752} {arXiv:1209.5752 [astro-ph.CO]} \BibitemShut
  {NoStop}%
\bibitem [{\citenamefont {Vogelsberger}\ \emph {et~al.}(2012)\citenamefont
  {Vogelsberger}, \citenamefont {Zavala},\ and\ \citenamefont
  {Loeb}}]{Vogelsberger:2012ku}%
  \BibitemOpen
  \bibfield  {author} {\bibinfo {author} {\bibfnamefont {M.}~\bibnamefont
  {Vogelsberger}}, \bibinfo {author} {\bibfnamefont {J.}~\bibnamefont
  {Zavala}}, \ and\ \bibinfo {author} {\bibfnamefont {A.}~\bibnamefont
  {Loeb}},\ }\href {\doibase 10.1111/j.1365-2966.2012.21182.x} {\bibfield
  {journal} {\bibinfo  {journal} {Mon. Not. Roy. Astron. Soc.}\ }\textbf
  {\bibinfo {volume} {423}},\ \bibinfo {pages} {3740} (\bibinfo {year}
  {2012})},\ \Eprint {http://arxiv.org/abs/1201.5892} {arXiv:1201.5892
  [astro-ph.CO]} \BibitemShut {NoStop}%
\bibitem [{\citenamefont {Vogelsberger}\ \emph {et~al.}(2016)\citenamefont
  {Vogelsberger}, \citenamefont {Zavala}, \citenamefont {Cyr-Racine},
  \citenamefont {Pfrommer}, \citenamefont {Bringmann},\ and\ \citenamefont
  {Sigurdson}}]{Vogelsberger:2015gpr}%
  \BibitemOpen
  \bibfield  {author} {\bibinfo {author} {\bibfnamefont {M.}~\bibnamefont
  {Vogelsberger}}, \bibinfo {author} {\bibfnamefont {J.}~\bibnamefont
  {Zavala}}, \bibinfo {author} {\bibfnamefont {F.-Y.}\ \bibnamefont
  {Cyr-Racine}}, \bibinfo {author} {\bibfnamefont {C.}~\bibnamefont
  {Pfrommer}}, \bibinfo {author} {\bibfnamefont {T.}~\bibnamefont {Bringmann}},
  \ and\ \bibinfo {author} {\bibfnamefont {K.}~\bibnamefont {Sigurdson}},\
  }\href {\doibase 10.1093/mnras/stw1076} {\bibfield  {journal} {\bibinfo
  {journal} {Mon. Not. Roy. Astron. Soc.}\ }\textbf {\bibinfo {volume} {460}},\
  \bibinfo {pages} {1399} (\bibinfo {year} {2016})},\ \Eprint
  {http://arxiv.org/abs/1512.05349} {arXiv:1512.05349 [astro-ph.CO]}
  \BibitemShut {NoStop}%
\bibitem [{\citenamefont {Huo}\ \emph {et~al.}(2018)\citenamefont {Huo},
  \citenamefont {Kaplinghat}, \citenamefont {Pan},\ and\ \citenamefont
  {Yu}}]{Huo:2017vef}%
  \BibitemOpen
  \bibfield  {author} {\bibinfo {author} {\bibfnamefont {R.}~\bibnamefont
  {Huo}}, \bibinfo {author} {\bibfnamefont {M.}~\bibnamefont {Kaplinghat}},
  \bibinfo {author} {\bibfnamefont {Z.}~\bibnamefont {Pan}}, \ and\ \bibinfo
  {author} {\bibfnamefont {H.-B.}\ \bibnamefont {Yu}},\ }\href {\doibase
  10.1016/j.physletb.2018.06.024} {\bibfield  {journal} {\bibinfo  {journal}
  {Phys. Lett.}\ }\textbf {\bibinfo {volume} {B783}},\ \bibinfo {pages} {76}
  (\bibinfo {year} {2018})},\ \Eprint {http://arxiv.org/abs/1709.09717}
  {arXiv:1709.09717 [hep-ph]} \BibitemShut {NoStop}%
\bibitem [{\citenamefont {Hawking}(1971)}]{Hawking:1971ei}%
  \BibitemOpen
  \bibfield  {author} {\bibinfo {author} {\bibfnamefont {S.}~\bibnamefont
  {Hawking}},\ }\href@noop {} {\bibfield  {journal} {\bibinfo  {journal} {Mon.
  Not. Roy. Astron. Soc.}\ }\textbf {\bibinfo {volume} {152}},\ \bibinfo
  {pages} {75} (\bibinfo {year} {1971})}\BibitemShut {NoStop}%
\bibitem [{\citenamefont {Carr}\ and\ \citenamefont
  {Hawking}(1974)}]{Carr:1974nx}%
  \BibitemOpen
  \bibfield  {author} {\bibinfo {author} {\bibfnamefont {B.~J.}\ \bibnamefont
  {Carr}}\ and\ \bibinfo {author} {\bibfnamefont {S.~W.}\ \bibnamefont
  {Hawking}},\ }\href@noop {} {\bibfield  {journal} {\bibinfo  {journal} {Mon.
  Not. Roy. Astron. Soc.}\ }\textbf {\bibinfo {volume} {168}},\ \bibinfo
  {pages} {399} (\bibinfo {year} {1974})}\BibitemShut {NoStop}%
\bibitem [{\citenamefont {Carr}\ \emph {et~al.}(2016)\citenamefont {Carr},
  \citenamefont {Kuhnel},\ and\ \citenamefont {Sandstad}}]{Carr:2016drx}%
  \BibitemOpen
  \bibfield  {author} {\bibinfo {author} {\bibfnamefont {B.}~\bibnamefont
  {Carr}}, \bibinfo {author} {\bibfnamefont {F.}~\bibnamefont {Kuhnel}}, \ and\
  \bibinfo {author} {\bibfnamefont {M.}~\bibnamefont {Sandstad}},\ }\href
  {\doibase 10.1103/PhysRevD.94.083504} {\bibfield  {journal} {\bibinfo
  {journal} {Phys. Rev.}\ }\textbf {\bibinfo {volume} {D94}},\ \bibinfo {pages}
  {083504} (\bibinfo {year} {2016})},\ \Eprint
  {http://arxiv.org/abs/1607.06077} {arXiv:1607.06077 [astro-ph.CO]}
  \BibitemShut {NoStop}%
\bibitem [{\citenamefont {Murgia}\ \emph {et~al.}(2019)\citenamefont {Murgia},
  \citenamefont {Scelfo}, \citenamefont {Viel},\ and\ \citenamefont
  {Raccanelli}}]{Murgia:2019duy}%
  \BibitemOpen
  \bibfield  {author} {\bibinfo {author} {\bibfnamefont {R.}~\bibnamefont
  {Murgia}}, \bibinfo {author} {\bibfnamefont {G.}~\bibnamefont {Scelfo}},
  \bibinfo {author} {\bibfnamefont {M.}~\bibnamefont {Viel}}, \ and\ \bibinfo
  {author} {\bibfnamefont {A.}~\bibnamefont {Raccanelli}},\ }\href {\doibase
  10.1103/PhysRevLett.123.071102} {\bibfield  {journal} {\bibinfo  {journal}
  {Phys. Rev. Lett.}\ }\textbf {\bibinfo {volume} {123}},\ \bibinfo {pages}
  {071102} (\bibinfo {year} {2019})},\ \Eprint
  {http://arxiv.org/abs/1903.10509} {arXiv:1903.10509 [astro-ph.CO]}
  \BibitemShut {NoStop}%
\bibitem [{\citenamefont {H{\"u}tsi}\ \emph {et~al.}(2019)\citenamefont
  {H{\"u}tsi}, \citenamefont {Raidal},\ and\ \citenamefont
  {Veerm{\"a}e}}]{Hutsi:2019hlw}%
  \BibitemOpen
  \bibfield  {author} {\bibinfo {author} {\bibfnamefont {G.}~\bibnamefont
  {H{\"u}tsi}}, \bibinfo {author} {\bibfnamefont {M.}~\bibnamefont {Raidal}}, \
  and\ \bibinfo {author} {\bibfnamefont {H.}~\bibnamefont {Veerm{\"a}e}},\
  }\href {\doibase 10.1103/PhysRevD.100.083016} {\bibfield  {journal} {\bibinfo
   {journal} {Phys. Rev.}\ }\textbf {\bibinfo {volume} {D100}},\ \bibinfo
  {pages} {083016} (\bibinfo {year} {2019})},\ \Eprint
  {http://arxiv.org/abs/1907.06533} {arXiv:1907.06533 [astro-ph.CO]}
  \BibitemShut {NoStop}%
\bibitem [{\citenamefont {Inman}\ and\ \citenamefont
  {Ali-Ha{\"\i}moud}(2019)}]{Inman:2019wvr}%
  \BibitemOpen
  \bibfield  {author} {\bibinfo {author} {\bibfnamefont {D.}~\bibnamefont
  {Inman}}\ and\ \bibinfo {author} {\bibfnamefont {Y.}~\bibnamefont
  {Ali-Ha{\"\i}moud}},\ }\href {\doibase 10.1103/PhysRevD.100.083528}
  {\bibfield  {journal} {\bibinfo  {journal} {Phys. Rev.}\ }\textbf {\bibinfo
  {volume} {D100}},\ \bibinfo {pages} {083528} (\bibinfo {year} {2019})},\
  \Eprint {http://arxiv.org/abs/1907.08129} {arXiv:1907.08129 [astro-ph.CO]}
  \BibitemShut {NoStop}%
\bibitem [{\citenamefont {Vegetti}\ \emph {et~al.}(2010)\citenamefont
  {Vegetti}, \citenamefont {Koopmans}, \citenamefont {Bolton}, \citenamefont
  {Treu},\ and\ \citenamefont {Gavazzi}}]{Vegetti:2009cz}%
  \BibitemOpen
  \bibfield  {author} {\bibinfo {author} {\bibfnamefont {S.}~\bibnamefont
  {Vegetti}}, \bibinfo {author} {\bibfnamefont {L.~V.~E.}\ \bibnamefont
  {Koopmans}}, \bibinfo {author} {\bibfnamefont {A.}~\bibnamefont {Bolton}},
  \bibinfo {author} {\bibfnamefont {T.}~\bibnamefont {Treu}}, \ and\ \bibinfo
  {author} {\bibfnamefont {R.}~\bibnamefont {Gavazzi}},\ }\href {\doibase
  10.1111/j.1365-2966.2010.16865.x} {\bibfield  {journal} {\bibinfo  {journal}
  {Mon. Not. Roy. Astron. Soc.}\ }\textbf {\bibinfo {volume} {408}},\ \bibinfo
  {pages} {1969} (\bibinfo {year} {2010})},\ \Eprint
  {http://arxiv.org/abs/0910.0760} {arXiv:0910.0760 [astro-ph.CO]} \BibitemShut
  {NoStop}%
\bibitem [{\citenamefont {Vegetti}\ \emph {et~al.}(2012)\citenamefont
  {Vegetti}, \citenamefont {Lagattuta}, \citenamefont {McKean}, \citenamefont
  {Auger}, \citenamefont {Fassnacht},\ and\ \citenamefont
  {Koopmans}}]{Vegetti:2012mc}%
  \BibitemOpen
  \bibfield  {author} {\bibinfo {author} {\bibfnamefont {S.}~\bibnamefont
  {Vegetti}}, \bibinfo {author} {\bibfnamefont {D.~J.}\ \bibnamefont
  {Lagattuta}}, \bibinfo {author} {\bibfnamefont {J.~P.}\ \bibnamefont
  {McKean}}, \bibinfo {author} {\bibfnamefont {M.~W.}\ \bibnamefont {Auger}},
  \bibinfo {author} {\bibfnamefont {C.~D.}\ \bibnamefont {Fassnacht}}, \ and\
  \bibinfo {author} {\bibfnamefont {L.~V.~E.}\ \bibnamefont {Koopmans}},\
  }\href {\doibase 10.1038/nature10669} {\bibfield  {journal} {\bibinfo
  {journal} {Nature}\ }\textbf {\bibinfo {volume} {481}},\ \bibinfo {pages}
  {341} (\bibinfo {year} {2012})},\ \Eprint {http://arxiv.org/abs/1201.3643}
  {arXiv:1201.3643 [astro-ph.CO]} \BibitemShut {NoStop}%
\bibitem [{\citenamefont {Hezaveh}\ \emph
  {et~al.}(2016{\natexlab{a}})\citenamefont {Hezaveh}, \citenamefont {Dalal},
  \citenamefont {Holder}, \citenamefont {Kisner}, \citenamefont {Kuhlen},\ and\
  \citenamefont {Perreault~Levasseur}}]{Hezaveh:2014aoa}%
  \BibitemOpen
  \bibfield  {author} {\bibinfo {author} {\bibfnamefont {Y.}~\bibnamefont
  {Hezaveh}}, \bibinfo {author} {\bibfnamefont {N.}~\bibnamefont {Dalal}},
  \bibinfo {author} {\bibfnamefont {G.}~\bibnamefont {Holder}}, \bibinfo
  {author} {\bibfnamefont {T.}~\bibnamefont {Kisner}}, \bibinfo {author}
  {\bibfnamefont {M.}~\bibnamefont {Kuhlen}}, \ and\ \bibinfo {author}
  {\bibfnamefont {L.}~\bibnamefont {Perreault~Levasseur}},\ }\href {\doibase
  10.1088/1475-7516/2016/11/048} {\bibfield  {journal} {\bibinfo  {journal}
  {JCAP}\ }\textbf {\bibinfo {volume} {1611}},\ \bibinfo {pages} {048}
  (\bibinfo {year} {2016}{\natexlab{a}})},\ \Eprint
  {http://arxiv.org/abs/1403.2720} {arXiv:1403.2720 [astro-ph.CO]} \BibitemShut
  {NoStop}%
\bibitem [{\citenamefont {Hezaveh}\ \emph
  {et~al.}(2016{\natexlab{b}})\citenamefont {Hezaveh} \emph
  {et~al.}}]{Hezaveh:2016ltk}%
  \BibitemOpen
  \bibfield  {author} {\bibinfo {author} {\bibfnamefont {Y.~D.}\ \bibnamefont
  {Hezaveh}} \emph {et~al.},\ }\href {\doibase 10.3847/0004-637X/823/1/37}
  {\bibfield  {journal} {\bibinfo  {journal} {Astrophys. J.}\ }\textbf
  {\bibinfo {volume} {823}},\ \bibinfo {pages} {37} (\bibinfo {year}
  {2016}{\natexlab{b}})},\ \Eprint {http://arxiv.org/abs/1601.01388}
  {arXiv:1601.01388 [astro-ph.CO]} \BibitemShut {NoStop}%
\bibitem [{\citenamefont {{D{\'\i}az Rivero}}\ \emph
  {et~al.}(2018)\citenamefont {{D{\'\i}az Rivero}}, \citenamefont {{Dvorkin}},
  \citenamefont {{Cyr-Racine}}, \citenamefont {{Zavala}},\ and\ \citenamefont
  {{Vogelsberger}}}]{2018PhRvD..98j3517D}%
  \BibitemOpen
  \bibfield  {author} {\bibinfo {author} {\bibfnamefont {A.}~\bibnamefont
  {{D{\'\i}az Rivero}}}, \bibinfo {author} {\bibfnamefont {C.}~\bibnamefont
  {{Dvorkin}}}, \bibinfo {author} {\bibfnamefont {F.-Y.}\ \bibnamefont
  {{Cyr-Racine}}}, \bibinfo {author} {\bibfnamefont {J.}~\bibnamefont
  {{Zavala}}}, \ and\ \bibinfo {author} {\bibfnamefont {M.}~\bibnamefont
  {{Vogelsberger}}},\ }\href {\doibase 10.1103/PhysRevD.98.103517} {\bibfield
  {journal} {\bibinfo  {journal} {\prd}\ }\textbf {\bibinfo {volume} {98}},\
  \bibinfo {eid} {103517} (\bibinfo {year} {2018})},\ \Eprint
  {http://arxiv.org/abs/1809.00004} {arXiv:1809.00004 [astro-ph.CO]}
  \BibitemShut {NoStop}%
\bibitem [{\citenamefont {{Brehmer}}\ \emph {et~al.}(2019)\citenamefont
  {{Brehmer}}, \citenamefont {{Mishra-Sharma}}, \citenamefont {{Hermans}},
  \citenamefont {{Louppe}},\ and\ \citenamefont
  {{Cranmer}}}]{2019arXiv190902005B}%
  \BibitemOpen
  \bibfield  {author} {\bibinfo {author} {\bibfnamefont {J.}~\bibnamefont
  {{Brehmer}}}, \bibinfo {author} {\bibfnamefont {S.}~\bibnamefont
  {{Mishra-Sharma}}}, \bibinfo {author} {\bibfnamefont {J.}~\bibnamefont
  {{Hermans}}}, \bibinfo {author} {\bibfnamefont {G.}~\bibnamefont {{Louppe}}},
  \ and\ \bibinfo {author} {\bibfnamefont {K.}~\bibnamefont {{Cranmer}}},\
  }\href@noop {} {\bibfield  {journal} {\bibinfo  {journal} {arXiv e-prints}\
  ,\ \bibinfo {eid} {arXiv:1909.02005}} (\bibinfo {year} {2019})},\ \Eprint
  {http://arxiv.org/abs/1909.02005} {arXiv:1909.02005 [astro-ph.CO]}
  \BibitemShut {NoStop}%
\bibitem [{\citenamefont {Hsueh}\ \emph {et~al.}(2019)\citenamefont {Hsueh},
  \citenamefont {Enzi}, \citenamefont {Vegetti}, \citenamefont {Auger},
  \citenamefont {Fassnacht}, \citenamefont {Despali}, \citenamefont
  {Koopmans},\ and\ \citenamefont {McKean}}]{Hsueh:2019ynk}%
  \BibitemOpen
  \bibfield  {author} {\bibinfo {author} {\bibfnamefont {J.-W.}\ \bibnamefont
  {Hsueh}}, \bibinfo {author} {\bibfnamefont {W.}~\bibnamefont {Enzi}},
  \bibinfo {author} {\bibfnamefont {S.}~\bibnamefont {Vegetti}}, \bibinfo
  {author} {\bibfnamefont {M.}~\bibnamefont {Auger}}, \bibinfo {author}
  {\bibfnamefont {C.~D.}\ \bibnamefont {Fassnacht}}, \bibinfo {author}
  {\bibfnamefont {G.}~\bibnamefont {Despali}}, \bibinfo {author} {\bibfnamefont
  {L.~V.~E.}\ \bibnamefont {Koopmans}}, \ and\ \bibinfo {author} {\bibfnamefont
  {J.~P.}\ \bibnamefont {McKean}},\ }\href@noop {} {\  (\bibinfo {year}
  {2019})},\ \Eprint {http://arxiv.org/abs/1905.04182} {arXiv:1905.04182
  [astro-ph.CO]} \BibitemShut {NoStop}%
\bibitem [{\citenamefont {Alexander}\ \emph
  {et~al.}(2019{\natexlab{a}})\citenamefont {Alexander}, \citenamefont
  {Gleyzer}, \citenamefont {McDonough}, \citenamefont {Toomey},\ and\
  \citenamefont {Usai}}]{Alexander:2019puy}%
  \BibitemOpen
  \bibfield  {author} {\bibinfo {author} {\bibfnamefont {S.}~\bibnamefont
  {Alexander}}, \bibinfo {author} {\bibfnamefont {S.}~\bibnamefont {Gleyzer}},
  \bibinfo {author} {\bibfnamefont {E.}~\bibnamefont {McDonough}}, \bibinfo
  {author} {\bibfnamefont {M.~W.}\ \bibnamefont {Toomey}}, \ and\ \bibinfo
  {author} {\bibfnamefont {E.}~\bibnamefont {Usai}},\ }\href@noop {} {\
  (\bibinfo {year} {2019}{\natexlab{a}})},\ \Eprint
  {http://arxiv.org/abs/1909.07346} {arXiv:1909.07346 [astro-ph.CO]}
  \BibitemShut {NoStop}%
\bibitem [{\citenamefont {{Gilman}}\ \emph {et~al.}(2020)\citenamefont
  {{Gilman}}, \citenamefont {{Birrer}}, \citenamefont {{Nierenberg}},
  \citenamefont {{Treu}}, \citenamefont {{Du}},\ and\ \citenamefont
  {{Benson}}}]{2020MNRAS.491.6077G}%
  \BibitemOpen
  \bibfield  {author} {\bibinfo {author} {\bibfnamefont {D.}~\bibnamefont
  {{Gilman}}}, \bibinfo {author} {\bibfnamefont {S.}~\bibnamefont {{Birrer}}},
  \bibinfo {author} {\bibfnamefont {A.}~\bibnamefont {{Nierenberg}}}, \bibinfo
  {author} {\bibfnamefont {T.}~\bibnamefont {{Treu}}}, \bibinfo {author}
  {\bibfnamefont {X.}~\bibnamefont {{Du}}}, \ and\ \bibinfo {author}
  {\bibfnamefont {A.}~\bibnamefont {{Benson}}},\ }\href {\doibase
  10.1093/mnras/stz3480} {\bibfield  {journal} {\bibinfo  {journal} {\mnras}\
  }\textbf {\bibinfo {volume} {491}},\ \bibinfo {pages} {6077} (\bibinfo {year}
  {2020})},\ \Eprint {http://arxiv.org/abs/1908.06983} {arXiv:1908.06983
  [astro-ph.CO]} \BibitemShut {NoStop}%
\bibitem [{\citenamefont {Croft}\ \emph {et~al.}(1998)\citenamefont {Croft},
  \citenamefont {Weinberg}, \citenamefont {Katz},\ and\ \citenamefont
  {Hernquist}}]{Croft:1997jf}%
  \BibitemOpen
  \bibfield  {author} {\bibinfo {author} {\bibfnamefont {R.~A.~C.}\
  \bibnamefont {Croft}}, \bibinfo {author} {\bibfnamefont {D.~H.}\ \bibnamefont
  {Weinberg}}, \bibinfo {author} {\bibfnamefont {N.}~\bibnamefont {Katz}}, \
  and\ \bibinfo {author} {\bibfnamefont {L.}~\bibnamefont {Hernquist}},\ }\href
  {\doibase 10.1086/305289} {\bibfield  {journal} {\bibinfo  {journal}
  {Astrophys. J.}\ }\textbf {\bibinfo {volume} {495}},\ \bibinfo {pages} {44}
  (\bibinfo {year} {1998})},\ \Eprint {http://arxiv.org/abs/astro-ph/9708018}
  {arXiv:astro-ph/9708018 [astro-ph]} \BibitemShut {NoStop}%
\bibitem [{\citenamefont {Croft}\ \emph {et~al.}(2002)\citenamefont {Croft},
  \citenamefont {Weinberg}, \citenamefont {Bolte}, \citenamefont {Burles},
  \citenamefont {Hernquist}, \citenamefont {Katz}, \citenamefont {Kirkman},\
  and\ \citenamefont {Tytler}}]{Croft:2000hs}%
  \BibitemOpen
  \bibfield  {author} {\bibinfo {author} {\bibfnamefont {R.~A.~C.}\
  \bibnamefont {Croft}}, \bibinfo {author} {\bibfnamefont {D.~H.}\ \bibnamefont
  {Weinberg}}, \bibinfo {author} {\bibfnamefont {M.}~\bibnamefont {Bolte}},
  \bibinfo {author} {\bibfnamefont {S.}~\bibnamefont {Burles}}, \bibinfo
  {author} {\bibfnamefont {L.}~\bibnamefont {Hernquist}}, \bibinfo {author}
  {\bibfnamefont {N.}~\bibnamefont {Katz}}, \bibinfo {author} {\bibfnamefont
  {D.}~\bibnamefont {Kirkman}}, \ and\ \bibinfo {author} {\bibfnamefont
  {D.}~\bibnamefont {Tytler}},\ }\href {\doibase 10.1086/344099} {\bibfield
  {journal} {\bibinfo  {journal} {Astrophys. J.}\ }\textbf {\bibinfo {volume}
  {581}},\ \bibinfo {pages} {20} (\bibinfo {year} {2002})},\ \Eprint
  {http://arxiv.org/abs/astro-ph/0012324} {arXiv:astro-ph/0012324 [astro-ph]}
  \BibitemShut {NoStop}%
\bibitem [{\citenamefont {Viel}\ \emph {et~al.}(2013)\citenamefont {Viel},
  \citenamefont {Becker}, \citenamefont {Bolton},\ and\ \citenamefont
  {Haehnelt}}]{Viel:2013apy}%
  \BibitemOpen
  \bibfield  {author} {\bibinfo {author} {\bibfnamefont {M.}~\bibnamefont
  {Viel}}, \bibinfo {author} {\bibfnamefont {G.~D.}\ \bibnamefont {Becker}},
  \bibinfo {author} {\bibfnamefont {J.~S.}\ \bibnamefont {Bolton}}, \ and\
  \bibinfo {author} {\bibfnamefont {M.~G.}\ \bibnamefont {Haehnelt}},\ }\href
  {\doibase 10.1103/PhysRevD.88.043502} {\bibfield  {journal} {\bibinfo
  {journal} {Phys. Rev.}\ }\textbf {\bibinfo {volume} {D88}},\ \bibinfo {pages}
  {043502} (\bibinfo {year} {2013})},\ \Eprint {http://arxiv.org/abs/1306.2314}
  {arXiv:1306.2314 [astro-ph.CO]} \BibitemShut {NoStop}%
\bibitem [{\citenamefont {Seljak}\ \emph {et~al.}(2006)\citenamefont {Seljak},
  \citenamefont {Slosar},\ and\ \citenamefont {McDonald}}]{Seljak:2006bg}%
  \BibitemOpen
  \bibfield  {author} {\bibinfo {author} {\bibfnamefont {U.}~\bibnamefont
  {Seljak}}, \bibinfo {author} {\bibfnamefont {A.}~\bibnamefont {Slosar}}, \
  and\ \bibinfo {author} {\bibfnamefont {P.}~\bibnamefont {McDonald}},\ }\href
  {\doibase 10.1088/1475-7516/2006/10/014} {\bibfield  {journal} {\bibinfo
  {journal} {JCAP}\ }\textbf {\bibinfo {volume} {0610}},\ \bibinfo {pages}
  {014} (\bibinfo {year} {2006})},\ \Eprint
  {http://arxiv.org/abs/astro-ph/0604335} {arXiv:astro-ph/0604335 [astro-ph]}
  \BibitemShut {NoStop}%
\bibitem [{\citenamefont {{Boyarsky}}\ \emph {et~al.}(2009)\citenamefont
  {{Boyarsky}}, \citenamefont {{Lesgourgues}}, \citenamefont {{Ruchayskiy}},\
  and\ \citenamefont {{Viel}}}]{2009JCAP...05..012B}%
  \BibitemOpen
  \bibfield  {author} {\bibinfo {author} {\bibfnamefont {A.}~\bibnamefont
  {{Boyarsky}}}, \bibinfo {author} {\bibfnamefont {J.}~\bibnamefont
  {{Lesgourgues}}}, \bibinfo {author} {\bibfnamefont {O.}~\bibnamefont
  {{Ruchayskiy}}}, \ and\ \bibinfo {author} {\bibfnamefont {M.}~\bibnamefont
  {{Viel}}},\ }\href {\doibase 10.1088/1475-7516/2009/05/012} {\bibfield
  {journal} {\bibinfo  {journal} {\jcap}\ }\textbf {\bibinfo {volume} {2009}},\
  \bibinfo {eid} {012} (\bibinfo {year} {2009})},\ \Eprint
  {http://arxiv.org/abs/0812.0010} {arXiv:0812.0010 [astro-ph]} \BibitemShut
  {NoStop}%
\bibitem [{\citenamefont {Ir{\v s}i{\v c}}\ \emph {et~al.}(2017)\citenamefont
  {Ir{\v s}i{\v c}}, \citenamefont {Viel}, \citenamefont {Haehnelt},
  \citenamefont {Bolton},\ and\ \citenamefont {Becker}}]{Irsic:2017yje}%
  \BibitemOpen
  \bibfield  {author} {\bibinfo {author} {\bibfnamefont {V.}~\bibnamefont
  {Ir{\v s}i{\v c}}}, \bibinfo {author} {\bibfnamefont {M.}~\bibnamefont
  {Viel}}, \bibinfo {author} {\bibfnamefont {M.~G.}\ \bibnamefont {Haehnelt}},
  \bibinfo {author} {\bibfnamefont {J.~S.}\ \bibnamefont {Bolton}}, \ and\
  \bibinfo {author} {\bibfnamefont {G.~D.}\ \bibnamefont {Becker}},\ }\href
  {\doibase 10.1103/PhysRevLett.119.031302} {\bibfield  {journal} {\bibinfo
  {journal} {Phys. Rev. Lett.}\ }\textbf {\bibinfo {volume} {119}},\ \bibinfo
  {pages} {031302} (\bibinfo {year} {2017})},\ \Eprint
  {http://arxiv.org/abs/1703.04683} {arXiv:1703.04683 [astro-ph.CO]}
  \BibitemShut {NoStop}%
\bibitem [{\citenamefont {{Palanque-Delabrouille}}\ \emph
  {et~al.}(2019)\citenamefont {{Palanque-Delabrouille}}, \citenamefont
  {{Y{\`e}che}}, \citenamefont {{Sch{\"o}neberg}}, \citenamefont
  {{Lesgourgues}}, \citenamefont {{Walther}}, \citenamefont {{Chabanier}},\
  and\ \citenamefont {{Armengaud}}}]{2019arXiv191109073P}%
  \BibitemOpen
  \bibfield  {author} {\bibinfo {author} {\bibfnamefont {N.}~\bibnamefont
  {{Palanque-Delabrouille}}}, \bibinfo {author} {\bibfnamefont
  {C.}~\bibnamefont {{Y{\`e}che}}}, \bibinfo {author} {\bibfnamefont
  {N.}~\bibnamefont {{Sch{\"o}neberg}}}, \bibinfo {author} {\bibfnamefont
  {J.}~\bibnamefont {{Lesgourgues}}}, \bibinfo {author} {\bibfnamefont
  {M.}~\bibnamefont {{Walther}}}, \bibinfo {author} {\bibfnamefont
  {S.}~\bibnamefont {{Chabanier}}}, \ and\ \bibinfo {author} {\bibfnamefont
  {E.}~\bibnamefont {{Armengaud}}},\ }\href@noop {} {\bibfield  {journal}
  {\bibinfo  {journal} {arXiv e-prints}\ ,\ \bibinfo {eid} {arXiv:1911.09073}}
  (\bibinfo {year} {2019})},\ \Eprint {http://arxiv.org/abs/1911.09073}
  {arXiv:1911.09073 [astro-ph.CO]} \BibitemShut {NoStop}%
\bibitem [{\citenamefont {{Ibata}}\ \emph {et~al.}(2002)\citenamefont
  {{Ibata}}, \citenamefont {{Lewis}}, \citenamefont {{Irwin}},\ and\
  \citenamefont {{Quinn}}}]{2002MNRAS.332..915I}%
  \BibitemOpen
  \bibfield  {author} {\bibinfo {author} {\bibfnamefont {R.~A.}\ \bibnamefont
  {{Ibata}}}, \bibinfo {author} {\bibfnamefont {G.~F.}\ \bibnamefont
  {{Lewis}}}, \bibinfo {author} {\bibfnamefont {M.~J.}\ \bibnamefont
  {{Irwin}}}, \ and\ \bibinfo {author} {\bibfnamefont {T.}~\bibnamefont
  {{Quinn}}},\ }\href {\doibase 10.1046/j.1365-8711.2002.05358.x} {\bibfield
  {journal} {\bibinfo  {journal} {\mnras}\ }\textbf {\bibinfo {volume} {332}},\
  \bibinfo {pages} {915} (\bibinfo {year} {2002})},\ \Eprint
  {http://arxiv.org/abs/astro-ph/0110690} {arXiv:astro-ph/0110690 [astro-ph]}
  \BibitemShut {NoStop}%
\bibitem [{\citenamefont {{Yoon}}\ \emph {et~al.}(2011)\citenamefont {{Yoon}},
  \citenamefont {{Johnston}},\ and\ \citenamefont {{Hogg}}}]{Yoon+2011}%
  \BibitemOpen
  \bibfield  {author} {\bibinfo {author} {\bibfnamefont {J.~H.}\ \bibnamefont
  {{Yoon}}}, \bibinfo {author} {\bibfnamefont {K.~V.}\ \bibnamefont
  {{Johnston}}}, \ and\ \bibinfo {author} {\bibfnamefont {D.~W.}\ \bibnamefont
  {{Hogg}}},\ }\href {\doibase 10.1088/0004-637X/731/1/58} {\bibfield
  {journal} {\bibinfo  {journal} {\apj}\ }\textbf {\bibinfo {volume} {731}},\
  \bibinfo {eid} {58} (\bibinfo {year} {2011})},\ \Eprint
  {http://arxiv.org/abs/1012.2884} {arXiv:1012.2884 [astro-ph.GA]} \BibitemShut
  {NoStop}%
\bibitem [{\citenamefont {{Carlberg}}(2012)}]{Carlberg2012}%
  \BibitemOpen
  \bibfield  {author} {\bibinfo {author} {\bibfnamefont {R.~G.}\ \bibnamefont
  {{Carlberg}}},\ }\href {\doibase 10.1088/0004-637X/748/1/20} {\bibfield
  {journal} {\bibinfo  {journal} {\apj}\ }\textbf {\bibinfo {volume} {748}},\
  \bibinfo {eid} {20} (\bibinfo {year} {2012})},\ \Eprint
  {http://arxiv.org/abs/1109.6022} {arXiv:1109.6022 [astro-ph.CO]} \BibitemShut
  {NoStop}%
\bibitem [{\citenamefont {Feldmann}\ and\ \citenamefont
  {Spolyar}(2015)}]{Feldmann:2013hqa}%
  \BibitemOpen
  \bibfield  {author} {\bibinfo {author} {\bibfnamefont {R.}~\bibnamefont
  {Feldmann}}\ and\ \bibinfo {author} {\bibfnamefont {D.}~\bibnamefont
  {Spolyar}},\ }\href {\doibase 10.1093/mnras/stu2147} {\bibfield  {journal}
  {\bibinfo  {journal} {Mon. Not. Roy. Astron. Soc.}\ }\textbf {\bibinfo
  {volume} {446}},\ \bibinfo {pages} {1000} (\bibinfo {year} {2015})},\ \Eprint
  {http://arxiv.org/abs/1310.2243} {arXiv:1310.2243 [astro-ph.GA]} \BibitemShut
  {NoStop}%
\bibitem [{\citenamefont {Bovy}\ \emph {et~al.}(2017)\citenamefont {Bovy},
  \citenamefont {Erkal},\ and\ \citenamefont {Sanders}}]{Bovy:2016irg}%
  \BibitemOpen
  \bibfield  {author} {\bibinfo {author} {\bibfnamefont {J.}~\bibnamefont
  {Bovy}}, \bibinfo {author} {\bibfnamefont {D.}~\bibnamefont {Erkal}}, \ and\
  \bibinfo {author} {\bibfnamefont {J.~L.}\ \bibnamefont {Sanders}},\ }\href
  {\doibase 10.1093/mnras/stw3067} {\bibfield  {journal} {\bibinfo  {journal}
  {Mon. Not. Roy. Astron. Soc.}\ }\textbf {\bibinfo {volume} {466}},\ \bibinfo
  {pages} {628} (\bibinfo {year} {2017})},\ \Eprint
  {http://arxiv.org/abs/1606.03470} {arXiv:1606.03470 [astro-ph.GA]}
  \BibitemShut {NoStop}%
\bibitem [{\citenamefont {{Erkal}}\ \emph {et~al.}(2016)\citenamefont
  {{Erkal}}, \citenamefont {{Belokurov}}, \citenamefont {{Bovy}},\ and\
  \citenamefont {{Sand ers}}}]{2016MNRAS.463..102E}%
  \BibitemOpen
  \bibfield  {author} {\bibinfo {author} {\bibfnamefont {D.}~\bibnamefont
  {{Erkal}}}, \bibinfo {author} {\bibfnamefont {V.}~\bibnamefont
  {{Belokurov}}}, \bibinfo {author} {\bibfnamefont {J.}~\bibnamefont {{Bovy}}},
  \ and\ \bibinfo {author} {\bibfnamefont {J.~L.}\ \bibnamefont {{Sand ers}}},\
  }\href {\doibase 10.1093/mnras/stw1957} {\bibfield  {journal} {\bibinfo
  {journal} {\mnras}\ }\textbf {\bibinfo {volume} {463}},\ \bibinfo {pages}
  {102} (\bibinfo {year} {2016})},\ \Eprint {http://arxiv.org/abs/1606.04946}
  {arXiv:1606.04946 [astro-ph.GA]} \BibitemShut {NoStop}%
\bibitem [{\citenamefont {Buschmann}\ \emph {et~al.}(2018)\citenamefont
  {Buschmann}, \citenamefont {Kopp}, \citenamefont {Safdi},\ and\ \citenamefont
  {Wu}}]{Buschmann:2017ams}%
  \BibitemOpen
  \bibfield  {author} {\bibinfo {author} {\bibfnamefont {M.}~\bibnamefont
  {Buschmann}}, \bibinfo {author} {\bibfnamefont {J.}~\bibnamefont {Kopp}},
  \bibinfo {author} {\bibfnamefont {B.~R.}\ \bibnamefont {Safdi}}, \ and\
  \bibinfo {author} {\bibfnamefont {C.-L.}\ \bibnamefont {Wu}},\ }\href
  {\doibase 10.1103/PhysRevLett.120.211101} {\bibfield  {journal} {\bibinfo
  {journal} {Phys. Rev. Lett.}\ }\textbf {\bibinfo {volume} {120}},\ \bibinfo
  {pages} {211101} (\bibinfo {year} {2018})},\ \Eprint
  {http://arxiv.org/abs/1711.03554} {arXiv:1711.03554 [astro-ph.GA]}
  \BibitemShut {NoStop}%
\bibitem [{\citenamefont {{Kim}}\ \emph {et~al.}(2017)\citenamefont {{Kim}},
  \citenamefont {{Peter}},\ and\ \citenamefont
  {{Hargis}}}]{2017arXiv171106267K}%
  \BibitemOpen
  \bibfield  {author} {\bibinfo {author} {\bibfnamefont {S.~Y.}\ \bibnamefont
  {{Kim}}}, \bibinfo {author} {\bibfnamefont {A.~H.~G.}\ \bibnamefont
  {{Peter}}}, \ and\ \bibinfo {author} {\bibfnamefont {J.~R.}\ \bibnamefont
  {{Hargis}}},\ }\href@noop {} {\bibfield  {journal} {\bibinfo  {journal}
  {arXiv e-prints}\ ,\ \bibinfo {eid} {arXiv:1711.06267}} (\bibinfo {year}
  {2017})},\ \Eprint {http://arxiv.org/abs/1711.06267} {arXiv:1711.06267
  [astro-ph.CO]} \BibitemShut {NoStop}%
\bibitem [{\citenamefont {{Escudero}}\ \emph {et~al.}(2018)\citenamefont
  {{Escudero}}, \citenamefont {{Lopez-Honorez}}, \citenamefont {{Mena}},
  \citenamefont {{Palomares-Ruiz}},\ and\ \citenamefont
  {{Villanueva-Domingo}}}]{2018JCAP...06..007E}%
  \BibitemOpen
  \bibfield  {author} {\bibinfo {author} {\bibfnamefont {M.}~\bibnamefont
  {{Escudero}}}, \bibinfo {author} {\bibfnamefont {L.}~\bibnamefont
  {{Lopez-Honorez}}}, \bibinfo {author} {\bibfnamefont {O.}~\bibnamefont
  {{Mena}}}, \bibinfo {author} {\bibfnamefont {S.}~\bibnamefont
  {{Palomares-Ruiz}}}, \ and\ \bibinfo {author} {\bibfnamefont
  {P.}~\bibnamefont {{Villanueva-Domingo}}},\ }\href {\doibase
  10.1088/1475-7516/2018/06/007} {\bibfield  {journal} {\bibinfo  {journal}
  {\jcap}\ }\textbf {\bibinfo {volume} {2018}},\ \bibinfo {eid} {007} (\bibinfo
  {year} {2018})},\ \Eprint {http://arxiv.org/abs/1803.08427} {arXiv:1803.08427
  [astro-ph.CO]} \BibitemShut {NoStop}%
\bibitem [{\citenamefont {{Jethwa}}\ \emph {et~al.}(2018)\citenamefont
  {{Jethwa}}, \citenamefont {{Erkal}},\ and\ \citenamefont
  {{Belokurov}}}]{2018MNRAS.473.2060J}%
  \BibitemOpen
  \bibfield  {author} {\bibinfo {author} {\bibfnamefont {P.}~\bibnamefont
  {{Jethwa}}}, \bibinfo {author} {\bibfnamefont {D.}~\bibnamefont {{Erkal}}}, \
  and\ \bibinfo {author} {\bibfnamefont {V.}~\bibnamefont {{Belokurov}}},\
  }\href {\doibase 10.1093/mnras/stx2330} {\bibfield  {journal} {\bibinfo
  {journal} {\mnras}\ }\textbf {\bibinfo {volume} {473}},\ \bibinfo {pages}
  {2060} (\bibinfo {year} {2018})},\ \Eprint {http://arxiv.org/abs/1612.07834}
  {arXiv:1612.07834 [astro-ph.GA]} \BibitemShut {NoStop}%
\bibitem [{\citenamefont {Marsh}\ and\ \citenamefont
  {Niemeyer}(2019)}]{Marsh:2018zyw}%
  \BibitemOpen
  \bibfield  {author} {\bibinfo {author} {\bibfnamefont {D.~J.~E.}\
  \bibnamefont {Marsh}}\ and\ \bibinfo {author} {\bibfnamefont {J.~C.}\
  \bibnamefont {Niemeyer}},\ }\href {\doibase 10.1103/PhysRevLett.123.051103}
  {\bibfield  {journal} {\bibinfo  {journal} {Phys. Rev. Lett.}\ }\textbf
  {\bibinfo {volume} {123}},\ \bibinfo {pages} {051103} (\bibinfo {year}
  {2019})},\ \Eprint {http://arxiv.org/abs/1810.08543} {arXiv:1810.08543
  [astro-ph.CO]} \BibitemShut {NoStop}%
\bibitem [{\citenamefont {Banik}\ \emph {et~al.}(2018)\citenamefont {Banik},
  \citenamefont {Bertone}, \citenamefont {Bovy},\ and\ \citenamefont
  {Bozorgnia}}]{Banik:2018pjp}%
  \BibitemOpen
  \bibfield  {author} {\bibinfo {author} {\bibfnamefont {N.}~\bibnamefont
  {Banik}}, \bibinfo {author} {\bibfnamefont {G.}~\bibnamefont {Bertone}},
  \bibinfo {author} {\bibfnamefont {J.}~\bibnamefont {Bovy}}, \ and\ \bibinfo
  {author} {\bibfnamefont {N.}~\bibnamefont {Bozorgnia}},\ }\href {\doibase
  10.1088/1475-7516/2018/07/061} {\bibfield  {journal} {\bibinfo  {journal}
  {JCAP}\ }\textbf {\bibinfo {volume} {1807}},\ \bibinfo {pages} {061}
  (\bibinfo {year} {2018})},\ \Eprint {http://arxiv.org/abs/1804.04384}
  {arXiv:1804.04384 [astro-ph.CO]} \BibitemShut {NoStop}%
\bibitem [{\citenamefont {Schive}\ \emph {et~al.}(2019)\citenamefont {Schive},
  \citenamefont {Chiueh},\ and\ \citenamefont {Broadhurst}}]{Schive:2019rrw}%
  \BibitemOpen
  \bibfield  {author} {\bibinfo {author} {\bibfnamefont {H.-Y.}\ \bibnamefont
  {Schive}}, \bibinfo {author} {\bibfnamefont {T.}~\bibnamefont {Chiueh}}, \
  and\ \bibinfo {author} {\bibfnamefont {T.}~\bibnamefont {Broadhurst}},\
  }\href@noop {} {\  (\bibinfo {year} {2019})},\ \Eprint
  {http://arxiv.org/abs/1912.09483} {arXiv:1912.09483 [astro-ph.GA]}
  \BibitemShut {NoStop}%
\bibitem [{\citenamefont {{Nadler}}\ \emph {et~al.}(2019)\citenamefont
  {{Nadler}}, \citenamefont {{Gluscevic}}, \citenamefont {{Boddy}},\ and\
  \citenamefont {{Wechsler}}}]{2019ApJ...878L..32N}%
  \BibitemOpen
  \bibfield  {author} {\bibinfo {author} {\bibfnamefont {E.~O.}\ \bibnamefont
  {{Nadler}}}, \bibinfo {author} {\bibfnamefont {V.}~\bibnamefont
  {{Gluscevic}}}, \bibinfo {author} {\bibfnamefont {K.~K.}\ \bibnamefont
  {{Boddy}}}, \ and\ \bibinfo {author} {\bibfnamefont {R.~H.}\ \bibnamefont
  {{Wechsler}}},\ }\href {\doibase 10.3847/2041-8213/ab1eb2} {\bibfield
  {journal} {\bibinfo  {journal} {\apjl}\ }\textbf {\bibinfo {volume} {878}},\
  \bibinfo {eid} {L32} (\bibinfo {year} {2019})},\ \Eprint
  {http://arxiv.org/abs/1904.10000} {arXiv:1904.10000 [astro-ph.CO]}
  \BibitemShut {NoStop}%
\bibitem [{\citenamefont {{Bonaca}}\ \emph {et~al.}(2019)\citenamefont
  {{Bonaca}}, \citenamefont {{Hogg}}, \citenamefont {{Price-Whelan}},\ and\
  \citenamefont {{Conroy}}}]{2019ApJ...880...38B}%
  \BibitemOpen
  \bibfield  {author} {\bibinfo {author} {\bibfnamefont {A.}~\bibnamefont
  {{Bonaca}}}, \bibinfo {author} {\bibfnamefont {D.~W.}\ \bibnamefont
  {{Hogg}}}, \bibinfo {author} {\bibfnamefont {A.~M.}\ \bibnamefont
  {{Price-Whelan}}}, \ and\ \bibinfo {author} {\bibfnamefont {C.}~\bibnamefont
  {{Conroy}}},\ }\href {\doibase 10.3847/1538-4357/ab2873} {\bibfield
  {journal} {\bibinfo  {journal} {\apj}\ }\textbf {\bibinfo {volume} {880}},\
  \bibinfo {eid} {38} (\bibinfo {year} {2019})},\ \Eprint
  {http://arxiv.org/abs/1811.03631} {arXiv:1811.03631 [astro-ph.GA]}
  \BibitemShut {NoStop}%
\bibitem [{\citenamefont {Banik}\ \emph
  {et~al.}(2019{\natexlab{a}})\citenamefont {Banik}, \citenamefont {Bovy},
  \citenamefont {Bertone}, \citenamefont {Erkal},\ and\ \citenamefont
  {de~Boer}}]{Banik:2019cza}%
  \BibitemOpen
  \bibfield  {author} {\bibinfo {author} {\bibfnamefont {N.}~\bibnamefont
  {Banik}}, \bibinfo {author} {\bibfnamefont {J.}~\bibnamefont {Bovy}},
  \bibinfo {author} {\bibfnamefont {G.}~\bibnamefont {Bertone}}, \bibinfo
  {author} {\bibfnamefont {D.}~\bibnamefont {Erkal}}, \ and\ \bibinfo {author}
  {\bibfnamefont {T.~J.~L.}\ \bibnamefont {de~Boer}},\ }\href@noop {} {\
  (\bibinfo {year} {2019}{\natexlab{a}})},\ \Eprint
  {http://arxiv.org/abs/1911.02662} {arXiv:1911.02662 [astro-ph.GA]}
  \BibitemShut {NoStop}%
\bibitem [{\citenamefont {Banik}\ \emph
  {et~al.}(2019{\natexlab{b}})\citenamefont {Banik}, \citenamefont {Bovy},
  \citenamefont {Bertone}, \citenamefont {Erkal},\ and\ \citenamefont
  {de~Boer}}]{Banik:2019smi}%
  \BibitemOpen
  \bibfield  {author} {\bibinfo {author} {\bibfnamefont {N.}~\bibnamefont
  {Banik}}, \bibinfo {author} {\bibfnamefont {J.}~\bibnamefont {Bovy}},
  \bibinfo {author} {\bibfnamefont {G.}~\bibnamefont {Bertone}}, \bibinfo
  {author} {\bibfnamefont {D.}~\bibnamefont {Erkal}}, \ and\ \bibinfo {author}
  {\bibfnamefont {T.~J.~L.}\ \bibnamefont {de~Boer}},\ }\href@noop {} {\
  (\bibinfo {year} {2019}{\natexlab{b}})},\ \Eprint
  {http://arxiv.org/abs/1911.02663} {arXiv:1911.02663 [astro-ph.GA]}
  \BibitemShut {NoStop}%
\bibitem [{\citenamefont {{Schneider}}\ \emph {et~al.}(2012)\citenamefont
  {{Schneider}}, \citenamefont {{Smith}}, \citenamefont {{Macci{\`o}}},\ and\
  \citenamefont {{Moore}}}]{Schneider+2012}%
  \BibitemOpen
  \bibfield  {author} {\bibinfo {author} {\bibfnamefont {A.}~\bibnamefont
  {{Schneider}}}, \bibinfo {author} {\bibfnamefont {R.~E.}\ \bibnamefont
  {{Smith}}}, \bibinfo {author} {\bibfnamefont {A.~V.}\ \bibnamefont
  {{Macci{\`o}}}}, \ and\ \bibinfo {author} {\bibfnamefont {B.}~\bibnamefont
  {{Moore}}},\ }\href {\doibase 10.1111/j.1365-2966.2012.21252.x} {\bibfield
  {journal} {\bibinfo  {journal} {\mnras}\ }\textbf {\bibinfo {volume} {424}},\
  \bibinfo {pages} {684} (\bibinfo {year} {2012})},\ \Eprint
  {http://arxiv.org/abs/1112.0330} {arXiv:1112.0330 [astro-ph.CO]} \BibitemShut
  {NoStop}%
\bibitem [{\citenamefont {{D'Onghia}}\ \emph {et~al.}(2010)\citenamefont
  {{D'Onghia}}, \citenamefont {{Springel}}, \citenamefont {{Hernquist}},\ and\
  \citenamefont {{Keres}}}]{Onguia+2009}%
  \BibitemOpen
  \bibfield  {author} {\bibinfo {author} {\bibfnamefont {E.}~\bibnamefont
  {{D'Onghia}}}, \bibinfo {author} {\bibfnamefont {V.}~\bibnamefont
  {{Springel}}}, \bibinfo {author} {\bibfnamefont {L.}~\bibnamefont
  {{Hernquist}}}, \ and\ \bibinfo {author} {\bibfnamefont {D.}~\bibnamefont
  {{Keres}}},\ }\href {\doibase 10.1088/0004-637X/709/2/1138} {\bibfield
  {journal} {\bibinfo  {journal} {\apj}\ }\textbf {\bibinfo {volume} {709}},\
  \bibinfo {pages} {1138} (\bibinfo {year} {2010})},\ \Eprint
  {http://arxiv.org/abs/0907.3482} {arXiv:0907.3482 [astro-ph.CO]} \BibitemShut
  {NoStop}%
\bibitem [{\citenamefont {{Sawala}}\ \emph {et~al.}(2017)\citenamefont
  {{Sawala}}, \citenamefont {{Pihajoki}}, \citenamefont {{Johansson}},
  \citenamefont {{Frenk}}, \citenamefont {{Navarro}}, \citenamefont {{Oman}},\
  and\ \citenamefont {{White}}}]{Sawala+2017}%
  \BibitemOpen
  \bibfield  {author} {\bibinfo {author} {\bibfnamefont {T.}~\bibnamefont
  {{Sawala}}}, \bibinfo {author} {\bibfnamefont {P.}~\bibnamefont
  {{Pihajoki}}}, \bibinfo {author} {\bibfnamefont {P.~H.}\ \bibnamefont
  {{Johansson}}}, \bibinfo {author} {\bibfnamefont {C.~S.}\ \bibnamefont
  {{Frenk}}}, \bibinfo {author} {\bibfnamefont {J.~F.}\ \bibnamefont
  {{Navarro}}}, \bibinfo {author} {\bibfnamefont {K.~A.}\ \bibnamefont
  {{Oman}}}, \ and\ \bibinfo {author} {\bibfnamefont {S.~D.~M.}\ \bibnamefont
  {{White}}},\ }\href {\doibase 10.1093/mnras/stx360} {\bibfield  {journal}
  {\bibinfo  {journal} {\mnras}\ }\textbf {\bibinfo {volume} {467}},\ \bibinfo
  {pages} {4383} (\bibinfo {year} {2017})},\ \Eprint
  {http://arxiv.org/abs/1609.01718} {arXiv:1609.01718 [astro-ph.GA]}
  \BibitemShut {NoStop}%
\bibitem [{\citenamefont {{Garrison-Kimmel}}\ \emph {et~al.}(2017)\citenamefont
  {{Garrison-Kimmel}}, \citenamefont {{Wetzel}}, \citenamefont {{Bullock}},
  \citenamefont {{Hopkins}}, \citenamefont {{Boylan-Kolchin}}, \citenamefont
  {{Faucher-Gigu{\`e}re}}, \citenamefont {{Kere{\v{s}}}}, \citenamefont
  {{Quataert}}, \citenamefont {{Sanderson}}, \citenamefont {{Graus}},\ and\
  \citenamefont {{Kelley}}}]{GarrisonKimmel+2017}%
  \BibitemOpen
  \bibfield  {author} {\bibinfo {author} {\bibfnamefont {S.}~\bibnamefont
  {{Garrison-Kimmel}}}, \bibinfo {author} {\bibfnamefont {A.}~\bibnamefont
  {{Wetzel}}}, \bibinfo {author} {\bibfnamefont {J.~S.}\ \bibnamefont
  {{Bullock}}}, \bibinfo {author} {\bibfnamefont {P.~F.}\ \bibnamefont
  {{Hopkins}}}, \bibinfo {author} {\bibfnamefont {M.}~\bibnamefont
  {{Boylan-Kolchin}}}, \bibinfo {author} {\bibfnamefont {C.-A.}\ \bibnamefont
  {{Faucher-Gigu{\`e}re}}}, \bibinfo {author} {\bibfnamefont {D.}~\bibnamefont
  {{Kere{\v{s}}}}}, \bibinfo {author} {\bibfnamefont {E.}~\bibnamefont
  {{Quataert}}}, \bibinfo {author} {\bibfnamefont {R.~E.}\ \bibnamefont
  {{Sanderson}}}, \bibinfo {author} {\bibfnamefont {A.~S.}\ \bibnamefont
  {{Graus}}}, \ and\ \bibinfo {author} {\bibfnamefont {T.}~\bibnamefont
  {{Kelley}}},\ }\href {\doibase 10.1093/mnras/stx1710} {\bibfield  {journal}
  {\bibinfo  {journal} {\mnras}\ }\textbf {\bibinfo {volume} {471}},\ \bibinfo
  {pages} {1709} (\bibinfo {year} {2017})},\ \Eprint
  {http://arxiv.org/abs/1701.03792} {arXiv:1701.03792 [astro-ph.GA]}
  \BibitemShut {NoStop}%
\bibitem [{\citenamefont {{Errani}}\ and\ \citenamefont
  {{Pe{\~n}arrubia}}(2019)}]{Errani&Penarrubia2019}%
  \BibitemOpen
  \bibfield  {author} {\bibinfo {author} {\bibfnamefont {R.}~\bibnamefont
  {{Errani}}}\ and\ \bibinfo {author} {\bibfnamefont {J.}~\bibnamefont
  {{Pe{\~n}arrubia}}},\ }\href@noop {} {\bibfield  {journal} {\bibinfo
  {journal} {arXiv e-prints}\ ,\ \bibinfo {eid} {arXiv:1906.01642}} (\bibinfo
  {year} {2019})},\ \Eprint {http://arxiv.org/abs/1906.01642} {arXiv:1906.01642
  [astro-ph.GA]} \BibitemShut {NoStop}%
\bibitem [{\citenamefont {Lovell}\ \emph {et~al.}(2014)\citenamefont {Lovell},
  \citenamefont {Frenk}, \citenamefont {Eke}, \citenamefont {Jenkins},
  \citenamefont {Gao},\ and\ \citenamefont {Theuns}}]{Lovell:2013ola}%
  \BibitemOpen
  \bibfield  {author} {\bibinfo {author} {\bibfnamefont {M.~R.}\ \bibnamefont
  {Lovell}}, \bibinfo {author} {\bibfnamefont {C.~S.}\ \bibnamefont {Frenk}},
  \bibinfo {author} {\bibfnamefont {V.~R.}\ \bibnamefont {Eke}}, \bibinfo
  {author} {\bibfnamefont {A.}~\bibnamefont {Jenkins}}, \bibinfo {author}
  {\bibfnamefont {L.}~\bibnamefont {Gao}}, \ and\ \bibinfo {author}
  {\bibfnamefont {T.}~\bibnamefont {Theuns}},\ }\href {\doibase
  10.1093/mnras/stt2431} {\bibfield  {journal} {\bibinfo  {journal}
  {Mon.Not.Roy.Astron.Soc.}\ }\textbf {\bibinfo {volume} {439}},\ \bibinfo
  {pages} {300} (\bibinfo {year} {2014})},\ \Eprint
  {http://arxiv.org/abs/1308.1399} {arXiv:1308.1399 [astro-ph.CO]} \BibitemShut
  {NoStop}%
\bibitem [{\citenamefont {Schneider}(2015)}]{Schneider:2014rda}%
  \BibitemOpen
  \bibfield  {author} {\bibinfo {author} {\bibfnamefont {A.}~\bibnamefont
  {Schneider}},\ }\href {\doibase 10.1093/mnras/stv1169} {\bibfield  {journal}
  {\bibinfo  {journal} {Mon. Not. Roy. Astron. Soc.}\ }\textbf {\bibinfo
  {volume} {451}},\ \bibinfo {pages} {3117} (\bibinfo {year} {2015})},\ \Eprint
  {http://arxiv.org/abs/1412.2133} {arXiv:1412.2133 [astro-ph.CO]} \BibitemShut
  {NoStop}%
\bibitem [{\citenamefont {{Viel}}\ \emph {et~al.}(2005)\citenamefont {{Viel}},
  \citenamefont {{Lesgourgues}}, \citenamefont {{Haehnelt}}, \citenamefont
  {{Matarrese}},\ and\ \citenamefont {{Riotto}}}]{Viel+2005}%
  \BibitemOpen
  \bibfield  {author} {\bibinfo {author} {\bibfnamefont {M.}~\bibnamefont
  {{Viel}}}, \bibinfo {author} {\bibfnamefont {J.}~\bibnamefont
  {{Lesgourgues}}}, \bibinfo {author} {\bibfnamefont {M.~G.}\ \bibnamefont
  {{Haehnelt}}}, \bibinfo {author} {\bibfnamefont {S.}~\bibnamefont
  {{Matarrese}}}, \ and\ \bibinfo {author} {\bibfnamefont {A.}~\bibnamefont
  {{Riotto}}},\ }\href {\doibase 10.1103/PhysRevD.71.063534} {\bibfield
  {journal} {\bibinfo  {journal} {\prd}\ }\textbf {\bibinfo {volume} {71}},\
  \bibinfo {eid} {063534} (\bibinfo {year} {2005})},\ \Eprint
  {http://arxiv.org/abs/astro-ph/0501562} {arXiv:astro-ph/0501562 [astro-ph]}
  \BibitemShut {NoStop}%
\bibitem [{\citenamefont {Marsh}(2016{\natexlab{b}})}]{Marsh:2016vgj}%
  \BibitemOpen
  \bibfield  {author} {\bibinfo {author} {\bibfnamefont {D.~J.~E.}\
  \bibnamefont {Marsh}},\ }\href@noop {} {\  (\bibinfo {year}
  {2016}{\natexlab{b}})},\ \Eprint {http://arxiv.org/abs/1605.05973}
  {arXiv:1605.05973 [astro-ph.CO]} \BibitemShut {NoStop}%
\bibitem [{\citenamefont {Du}\ \emph {et~al.}(2018)\citenamefont {Du},
  \citenamefont {Schwabe}, \citenamefont {Niemeyer},\ and\ \citenamefont
  {B{\"u}rger}}]{Du:2018zrg}%
  \BibitemOpen
  \bibfield  {author} {\bibinfo {author} {\bibfnamefont {X.}~\bibnamefont
  {Du}}, \bibinfo {author} {\bibfnamefont {B.}~\bibnamefont {Schwabe}},
  \bibinfo {author} {\bibfnamefont {J.~C.}\ \bibnamefont {Niemeyer}}, \ and\
  \bibinfo {author} {\bibfnamefont {D.}~\bibnamefont {B{\"u}rger}},\ }\href
  {\doibase 10.1103/PhysRevD.97.063507} {\bibfield  {journal} {\bibinfo
  {journal} {Phys. Rev.}\ }\textbf {\bibinfo {volume} {D97}},\ \bibinfo {pages}
  {063507} (\bibinfo {year} {2018})},\ \Eprint
  {http://arxiv.org/abs/1801.04864} {arXiv:1801.04864 [astro-ph.GA]}
  \BibitemShut {NoStop}%
\bibitem [{\citenamefont {Alexander}\ \emph
  {et~al.}(2019{\natexlab{b}})\citenamefont {Alexander}, \citenamefont
  {Bramburger},\ and\ \citenamefont {McDonough}}]{Alexander:2019qsh}%
  \BibitemOpen
  \bibfield  {author} {\bibinfo {author} {\bibfnamefont {S.}~\bibnamefont
  {Alexander}}, \bibinfo {author} {\bibfnamefont {J.~J.}\ \bibnamefont
  {Bramburger}}, \ and\ \bibinfo {author} {\bibfnamefont {E.}~\bibnamefont
  {McDonough}},\ }\href {\doibase 10.1016/j.physletb.2019.134871} {\bibfield
  {journal} {\bibinfo  {journal} {Phys. Lett.}\ }\textbf {\bibinfo {volume}
  {B797}},\ \bibinfo {pages} {134871} (\bibinfo {year} {2019}{\natexlab{b}})},\
  \Eprint {http://arxiv.org/abs/1901.03694} {arXiv:1901.03694 [astro-ph.CO]}
  \BibitemShut {NoStop}%
\bibitem [{\citenamefont {Boehm}\ \emph {et~al.}(2002)\citenamefont {Boehm},
  \citenamefont {Riazuelo}, \citenamefont {Hansen},\ and\ \citenamefont
  {Schaeffer}}]{Boehm:2001hm}%
  \BibitemOpen
  \bibfield  {author} {\bibinfo {author} {\bibfnamefont {C.}~\bibnamefont
  {Boehm}}, \bibinfo {author} {\bibfnamefont {A.}~\bibnamefont {Riazuelo}},
  \bibinfo {author} {\bibfnamefont {S.~H.}\ \bibnamefont {Hansen}}, \ and\
  \bibinfo {author} {\bibfnamefont {R.}~\bibnamefont {Schaeffer}},\ }\href
  {\doibase 10.1103/PhysRevD.66.083505} {\bibfield  {journal} {\bibinfo
  {journal} {Phys. Rev.}\ }\textbf {\bibinfo {volume} {D66}},\ \bibinfo {pages}
  {083505} (\bibinfo {year} {2002})},\ \Eprint
  {http://arxiv.org/abs/astro-ph/0112522} {arXiv:astro-ph/0112522 [astro-ph]}
  \BibitemShut {NoStop}%
\bibitem [{\citenamefont {Buckley}\ \emph {et~al.}(2014)\citenamefont
  {Buckley}, \citenamefont {Zavala}, \citenamefont {Cyr-Racine}, \citenamefont
  {Sigurdson},\ and\ \citenamefont {Vogelsberger}}]{Buckley:2014hja}%
  \BibitemOpen
  \bibfield  {author} {\bibinfo {author} {\bibfnamefont {M.~R.}\ \bibnamefont
  {Buckley}}, \bibinfo {author} {\bibfnamefont {J.}~\bibnamefont {Zavala}},
  \bibinfo {author} {\bibfnamefont {F.-Y.}\ \bibnamefont {Cyr-Racine}},
  \bibinfo {author} {\bibfnamefont {K.}~\bibnamefont {Sigurdson}}, \ and\
  \bibinfo {author} {\bibfnamefont {M.}~\bibnamefont {Vogelsberger}},\ }\href
  {\doibase 10.1103/PhysRevD.90.043524} {\bibfield  {journal} {\bibinfo
  {journal} {Phys. Rev.}\ }\textbf {\bibinfo {volume} {D90}},\ \bibinfo {pages}
  {043524} (\bibinfo {year} {2014})},\ \Eprint {http://arxiv.org/abs/1405.2075}
  {arXiv:1405.2075 [astro-ph.CO]} \BibitemShut {NoStop}%
\bibitem [{\citenamefont {Boehm}\ \emph {et~al.}(2014)\citenamefont {Boehm},
  \citenamefont {Schewtschenko}, \citenamefont {Wilkinson}, \citenamefont
  {Baugh},\ and\ \citenamefont {Pascoli}}]{Boehm:2014vja}%
  \BibitemOpen
  \bibfield  {author} {\bibinfo {author} {\bibfnamefont {C.}~\bibnamefont
  {Boehm}}, \bibinfo {author} {\bibfnamefont {J.~A.}\ \bibnamefont
  {Schewtschenko}}, \bibinfo {author} {\bibfnamefont {R.~J.}\ \bibnamefont
  {Wilkinson}}, \bibinfo {author} {\bibfnamefont {C.~M.}\ \bibnamefont
  {Baugh}}, \ and\ \bibinfo {author} {\bibfnamefont {S.}~\bibnamefont
  {Pascoli}},\ }\href {\doibase 10.1093/mnrasl/slu115} {\bibfield  {journal}
  {\bibinfo  {journal} {Mon. Not. Roy. Astron. Soc.}\ }\textbf {\bibinfo
  {volume} {445}},\ \bibinfo {pages} {L31} (\bibinfo {year} {2014})},\ \Eprint
  {http://arxiv.org/abs/1404.7012} {arXiv:1404.7012 [astro-ph.CO]} \BibitemShut
  {NoStop}%
\bibitem [{\citenamefont {Vogelsberger}\ \emph {et~al.}(2014)\citenamefont
  {Vogelsberger}, \citenamefont {Zavala}, \citenamefont {Simpson},\ and\
  \citenamefont {Jenkins}}]{Vogelsberger:2014pda}%
  \BibitemOpen
  \bibfield  {author} {\bibinfo {author} {\bibfnamefont {M.}~\bibnamefont
  {Vogelsberger}}, \bibinfo {author} {\bibfnamefont {J.}~\bibnamefont
  {Zavala}}, \bibinfo {author} {\bibfnamefont {C.}~\bibnamefont {Simpson}}, \
  and\ \bibinfo {author} {\bibfnamefont {A.}~\bibnamefont {Jenkins}},\ }\href
  {\doibase 10.1093/mnras/stu1713} {\bibfield  {journal} {\bibinfo  {journal}
  {Mon. Not. Roy. Astron. Soc.}\ }\textbf {\bibinfo {volume} {444}},\ \bibinfo
  {pages} {3684} (\bibinfo {year} {2014})},\ \Eprint
  {http://arxiv.org/abs/1405.5216} {arXiv:1405.5216 [astro-ph.CO]} \BibitemShut
  {NoStop}%
\bibitem [{\citenamefont {Kaplan}\ \emph {et~al.}(2010)\citenamefont {Kaplan},
  \citenamefont {Krnjaic}, \citenamefont {Rehermann},\ and\ \citenamefont
  {Wells}}]{Kaplan:2009de}%
  \BibitemOpen
  \bibfield  {author} {\bibinfo {author} {\bibfnamefont {D.~E.}\ \bibnamefont
  {Kaplan}}, \bibinfo {author} {\bibfnamefont {G.~Z.}\ \bibnamefont {Krnjaic}},
  \bibinfo {author} {\bibfnamefont {K.~R.}\ \bibnamefont {Rehermann}}, \ and\
  \bibinfo {author} {\bibfnamefont {C.~M.}\ \bibnamefont {Wells}},\ }\href
  {\doibase 10.1088/1475-7516/2010/05/021} {\bibfield  {journal} {\bibinfo
  {journal} {JCAP}\ }\textbf {\bibinfo {volume} {1005}},\ \bibinfo {pages}
  {021} (\bibinfo {year} {2010})},\ \Eprint {http://arxiv.org/abs/0909.0753}
  {arXiv:0909.0753 [hep-ph]} \BibitemShut {NoStop}%
\bibitem [{\citenamefont {{Johnston}}\ and\ \citenamefont
  {{Carlberg}}(2016)}]{2016ASSL..420..169J}%
  \BibitemOpen
  \bibfield  {author} {\bibinfo {author} {\bibfnamefont {K.~V.}\ \bibnamefont
  {{Johnston}}}\ and\ \bibinfo {author} {\bibfnamefont {R.~G.}\ \bibnamefont
  {{Carlberg}}},\ }\enquote {\bibinfo {title} {{Tidal Debris as a Dark Matter
  Probe}},}\ in\ \href {\doibase 10.1007/978-3-319-19336-6_7} {\emph {\bibinfo
  {booktitle} {Tidal Streams in the Local Group and Beyond, Astrophysics and
  Space Science Library, Volume 420. ISBN 978-3-319-19335-9. Springer
  International Publishing Switzerland, 2016, p. 169}}},\ \bibinfo {series}
  {Astrophysics and Space Science Library}, Vol.\ \bibinfo {volume} {420},\
  \bibinfo {editor} {edited by\ \bibinfo {editor} {\bibfnamefont {H.~J.}\
  \bibnamefont {{Newberg}}}\ and\ \bibinfo {editor} {\bibfnamefont {J.~L.}\
  \bibnamefont {{Carlin}}}}\ (\bibinfo {year} {2016})\ p.\ \bibinfo {pages}
  {169}\BibitemShut {NoStop}%
\bibitem [{\citenamefont {{Marsh}}\ and\ \citenamefont
  {{Pop}}(2015)}]{Marsh&Pop2015}%
  \BibitemOpen
  \bibfield  {author} {\bibinfo {author} {\bibfnamefont {D.~J.~E.}\
  \bibnamefont {{Marsh}}}\ and\ \bibinfo {author} {\bibfnamefont {A.-R.}\
  \bibnamefont {{Pop}}},\ }\href {\doibase 10.1093/mnras/stv1050} {\bibfield
  {journal} {\bibinfo  {journal} {\mnras}\ }\textbf {\bibinfo {volume} {451}},\
  \bibinfo {pages} {2479} (\bibinfo {year} {2015})},\ \Eprint
  {http://arxiv.org/abs/1502.03456} {arXiv:1502.03456 [astro-ph.CO]}
  \BibitemShut {NoStop}%
\bibitem [{\citenamefont {{Gonz{\'a}lez-Morales}}\ \emph
  {et~al.}(2017)\citenamefont {{Gonz{\'a}lez-Morales}}, \citenamefont
  {{Marsh}}, \citenamefont {{Pe{\~n}arrubia}},\ and\ \citenamefont
  {{Ure{\~n}a-L{\'o}pez}}}]{GonzalezMorales+17}%
  \BibitemOpen
  \bibfield  {author} {\bibinfo {author} {\bibfnamefont {A.~X.}\ \bibnamefont
  {{Gonz{\'a}lez-Morales}}}, \bibinfo {author} {\bibfnamefont {D.~J.~E.}\
  \bibnamefont {{Marsh}}}, \bibinfo {author} {\bibfnamefont {J.}~\bibnamefont
  {{Pe{\~n}arrubia}}}, \ and\ \bibinfo {author} {\bibfnamefont {L.~A.}\
  \bibnamefont {{Ure{\~n}a-L{\'o}pez}}},\ }\href {\doibase
  10.1093/mnras/stx1941} {\bibfield  {journal} {\bibinfo  {journal} {\mnras}\
  }\textbf {\bibinfo {volume} {472}},\ \bibinfo {pages} {1346} (\bibinfo {year}
  {2017})},\ \Eprint {http://arxiv.org/abs/1609.05856} {arXiv:1609.05856
  [astro-ph.CO]} \BibitemShut {NoStop}%
\bibitem [{\citenamefont {{Broadhurst}}\ \emph {et~al.}(2019)\citenamefont
  {{Broadhurst}}, \citenamefont {{de Martino}}, \citenamefont {{Nhan Luu}},
  \citenamefont {{Smoot}},\ and\ \citenamefont {{Tye}}}]{Broadhurst+19}%
  \BibitemOpen
  \bibfield  {author} {\bibinfo {author} {\bibfnamefont {T.}~\bibnamefont
  {{Broadhurst}}}, \bibinfo {author} {\bibfnamefont {I.}~\bibnamefont {{de
  Martino}}}, \bibinfo {author} {\bibfnamefont {H.}~\bibnamefont {{Nhan Luu}}},
  \bibinfo {author} {\bibfnamefont {G.~F.}\ \bibnamefont {{Smoot}}}, \ and\
  \bibinfo {author} {\bibfnamefont {S.~H.~H.}\ \bibnamefont {{Tye}}},\
  }\href@noop {} {\bibfield  {journal} {\bibinfo  {journal} {arXiv e-prints}\
  ,\ \bibinfo {eid} {arXiv:1902.10488}} (\bibinfo {year} {2019})},\ \Eprint
  {http://arxiv.org/abs/1902.10488} {arXiv:1902.10488 [astro-ph.CO]}
  \BibitemShut {NoStop}%
\bibitem [{\citenamefont {{Wasserman}}\ \emph {et~al.}(2019)\citenamefont
  {{Wasserman}}, \citenamefont {{van Dokkum}}, \citenamefont {{Romanowsky}},
  \citenamefont {{Brodie}}, \citenamefont {{Danieli}}, \citenamefont
  {{Forbes}}, \citenamefont {{Abraham}}, \citenamefont {{Martin}},
  \citenamefont {{Matuszewski}}, \citenamefont {{Villaume}}, \citenamefont
  {{Tamanas}},\ and\ \citenamefont {{Profumo}}}]{Wasserman+19}%
  \BibitemOpen
  \bibfield  {author} {\bibinfo {author} {\bibfnamefont {A.}~\bibnamefont
  {{Wasserman}}}, \bibinfo {author} {\bibfnamefont {P.}~\bibnamefont {{van
  Dokkum}}}, \bibinfo {author} {\bibfnamefont {A.~J.}\ \bibnamefont
  {{Romanowsky}}}, \bibinfo {author} {\bibfnamefont {J.}~\bibnamefont
  {{Brodie}}}, \bibinfo {author} {\bibfnamefont {S.}~\bibnamefont {{Danieli}}},
  \bibinfo {author} {\bibfnamefont {D.~A.}\ \bibnamefont {{Forbes}}}, \bibinfo
  {author} {\bibfnamefont {R.}~\bibnamefont {{Abraham}}}, \bibinfo {author}
  {\bibfnamefont {C.}~\bibnamefont {{Martin}}}, \bibinfo {author}
  {\bibfnamefont {M.}~\bibnamefont {{Matuszewski}}}, \bibinfo {author}
  {\bibfnamefont {A.}~\bibnamefont {{Villaume}}}, \bibinfo {author}
  {\bibfnamefont {J.}~\bibnamefont {{Tamanas}}}, \ and\ \bibinfo {author}
  {\bibfnamefont {S.}~\bibnamefont {{Profumo}}},\ }\href {\doibase
  10.3847/1538-4357/ab3eb9} {\bibfield  {journal} {\bibinfo  {journal} {\apj}\
  }\textbf {\bibinfo {volume} {885}},\ \bibinfo {eid} {155} (\bibinfo {year}
  {2019})},\ \Eprint {http://arxiv.org/abs/1905.10373} {arXiv:1905.10373
  [astro-ph.GA]} \BibitemShut {NoStop}%
\bibitem [{\citenamefont {{Oman}}\ \emph {et~al.}(2016)\citenamefont {{Oman}},
  \citenamefont {{Navarro}}, \citenamefont {{Sales}}, \citenamefont
  {{Fattahi}}, \citenamefont {{Frenk}}, \citenamefont {{Sawala}}, \citenamefont
  {{Schaller}},\ and\ \citenamefont {{White}}}]{Oman+16a}%
  \BibitemOpen
  \bibfield  {author} {\bibinfo {author} {\bibfnamefont {K.~A.}\ \bibnamefont
  {{Oman}}}, \bibinfo {author} {\bibfnamefont {J.~F.}\ \bibnamefont
  {{Navarro}}}, \bibinfo {author} {\bibfnamefont {L.~V.}\ \bibnamefont
  {{Sales}}}, \bibinfo {author} {\bibfnamefont {A.}~\bibnamefont {{Fattahi}}},
  \bibinfo {author} {\bibfnamefont {C.~S.}\ \bibnamefont {{Frenk}}}, \bibinfo
  {author} {\bibfnamefont {T.}~\bibnamefont {{Sawala}}}, \bibinfo {author}
  {\bibfnamefont {M.}~\bibnamefont {{Schaller}}}, \ and\ \bibinfo {author}
  {\bibfnamefont {S.~D.~M.}\ \bibnamefont {{White}}},\ }\href {\doibase
  10.1093/mnras/stw1251} {\bibfield  {journal} {\bibinfo  {journal} {\mnras}\
  }\textbf {\bibinfo {volume} {460}},\ \bibinfo {pages} {3610} (\bibinfo {year}
  {2016})},\ \Eprint {http://arxiv.org/abs/1601.01026} {arXiv:1601.01026
  [astro-ph.GA]} \BibitemShut {NoStop}%
\bibitem [{\citenamefont {{Ir{\v{s}}i{\v{c}}}}\ \emph
  {et~al.}(2017)\citenamefont {{Ir{\v{s}}i{\v{c}}}}, \citenamefont {{Viel}},
  \citenamefont {{Haehnelt}}, \citenamefont {{Bolton}},\ and\ \citenamefont
  {{Becker}}}]{2017PhRvL.119c1302I}%
  \BibitemOpen
  \bibfield  {author} {\bibinfo {author} {\bibfnamefont {V.}~\bibnamefont
  {{Ir{\v{s}}i{\v{c}}}}}, \bibinfo {author} {\bibfnamefont {M.}~\bibnamefont
  {{Viel}}}, \bibinfo {author} {\bibfnamefont {M.~G.}\ \bibnamefont
  {{Haehnelt}}}, \bibinfo {author} {\bibfnamefont {J.~S.}\ \bibnamefont
  {{Bolton}}}, \ and\ \bibinfo {author} {\bibfnamefont {G.~D.}\ \bibnamefont
  {{Becker}}},\ }\href {\doibase 10.1103/PhysRevLett.119.031302} {\bibfield
  {journal} {\bibinfo  {journal} {\prl}\ }\textbf {\bibinfo {volume} {119}},\
  \bibinfo {eid} {031302} (\bibinfo {year} {2017})},\ \Eprint
  {http://arxiv.org/abs/1703.04683} {arXiv:1703.04683 [astro-ph.CO]}
  \BibitemShut {NoStop}%
\bibitem [{\citenamefont {{Kobayashi}}\ \emph {et~al.}(2017)\citenamefont
  {{Kobayashi}}, \citenamefont {{Murgia}}, \citenamefont {{De Simone}},
  \citenamefont {{Ir{\v{s}}i{\v{c}}}},\ and\ \citenamefont
  {{Viel}}}]{2017PhRvD..96l3514K}%
  \BibitemOpen
  \bibfield  {author} {\bibinfo {author} {\bibfnamefont {T.}~\bibnamefont
  {{Kobayashi}}}, \bibinfo {author} {\bibfnamefont {R.}~\bibnamefont
  {{Murgia}}}, \bibinfo {author} {\bibfnamefont {A.}~\bibnamefont {{De
  Simone}}}, \bibinfo {author} {\bibfnamefont {V.}~\bibnamefont
  {{Ir{\v{s}}i{\v{c}}}}}, \ and\ \bibinfo {author} {\bibfnamefont
  {M.}~\bibnamefont {{Viel}}},\ }\href {\doibase 10.1103/PhysRevD.96.123514}
  {\bibfield  {journal} {\bibinfo  {journal} {\prd}\ }\textbf {\bibinfo
  {volume} {96}},\ \bibinfo {eid} {123514} (\bibinfo {year} {2017})},\ \Eprint
  {http://arxiv.org/abs/1708.00015} {arXiv:1708.00015 [astro-ph.CO]}
  \BibitemShut {NoStop}%
\bibitem [{\citenamefont {{Hlo{\v{z}}ek}}\ \emph {et~al.}(2018)\citenamefont
  {{Hlo{\v{z}}ek}}, \citenamefont {{Marsh}},\ and\ \citenamefont
  {{Grin}}}]{2018MNRAS.476.3063H}%
  \BibitemOpen
  \bibfield  {author} {\bibinfo {author} {\bibfnamefont {R.}~\bibnamefont
  {{Hlo{\v{z}}ek}}}, \bibinfo {author} {\bibfnamefont {D.~J.~E.}\ \bibnamefont
  {{Marsh}}}, \ and\ \bibinfo {author} {\bibfnamefont {D.}~\bibnamefont
  {{Grin}}},\ }\href {\doibase 10.1093/mnras/sty271} {\bibfield  {journal}
  {\bibinfo  {journal} {\mnras}\ }\textbf {\bibinfo {volume} {476}},\ \bibinfo
  {pages} {3063} (\bibinfo {year} {2018})},\ \Eprint
  {http://arxiv.org/abs/1708.05681} {arXiv:1708.05681 [astro-ph.CO]}
  \BibitemShut {NoStop}%
\bibitem [{\citenamefont {{Maleki}}\ \emph {et~al.}(2020)\citenamefont
  {{Maleki}}, \citenamefont {{Baghram}},\ and\ \citenamefont
  {{Rahvar}}}]{2020PhRvD.101b3508M}%
  \BibitemOpen
  \bibfield  {author} {\bibinfo {author} {\bibfnamefont {A.}~\bibnamefont
  {{Maleki}}}, \bibinfo {author} {\bibfnamefont {S.}~\bibnamefont {{Baghram}}},
  \ and\ \bibinfo {author} {\bibfnamefont {S.}~\bibnamefont {{Rahvar}}},\
  }\href {\doibase 10.1103/PhysRevD.101.023508} {\bibfield  {journal} {\bibinfo
   {journal} {\prd}\ }\textbf {\bibinfo {volume} {101}},\ \bibinfo {eid}
  {023508} (\bibinfo {year} {2020})},\ \Eprint
  {http://arxiv.org/abs/1911.00486} {arXiv:1911.00486 [astro-ph.CO]}
  \BibitemShut {NoStop}%
\bibitem [{\citenamefont {{Pasquini}}\ \emph {et~al.}(2004)\citenamefont
  {{Pasquini}}, \citenamefont {{Bonifacio}}, \citenamefont {{Randich}},
  \citenamefont {{Galli}},\ and\ \citenamefont
  {{Gratton}}}]{2004A&A...426..651P}%
  \BibitemOpen
  \bibfield  {author} {\bibinfo {author} {\bibfnamefont {L.}~\bibnamefont
  {{Pasquini}}}, \bibinfo {author} {\bibfnamefont {P.}~\bibnamefont
  {{Bonifacio}}}, \bibinfo {author} {\bibfnamefont {S.}~\bibnamefont
  {{Randich}}}, \bibinfo {author} {\bibfnamefont {D.}~\bibnamefont {{Galli}}},
  \ and\ \bibinfo {author} {\bibfnamefont {R.~G.}\ \bibnamefont {{Gratton}}},\
  }\href {\doibase 10.1051/0004-6361:20041254} {\bibfield  {journal} {\bibinfo
  {journal} {\aap}\ }\textbf {\bibinfo {volume} {426}},\ \bibinfo {pages} {651}
  (\bibinfo {year} {2004})},\ \Eprint {http://arxiv.org/abs/astro-ph/0407524}
  {arXiv:astro-ph/0407524 [astro-ph]} \BibitemShut {NoStop}%
\bibitem [{\citenamefont {{Eilers}}\ \emph {et~al.}(2019)\citenamefont
  {{Eilers}}, \citenamefont {{Hogg}}, \citenamefont {{Rix}},\ and\
  \citenamefont {{Ness}}}]{Eilers+2019}%
  \BibitemOpen
  \bibfield  {author} {\bibinfo {author} {\bibfnamefont {A.-C.}\ \bibnamefont
  {{Eilers}}}, \bibinfo {author} {\bibfnamefont {D.~W.}\ \bibnamefont
  {{Hogg}}}, \bibinfo {author} {\bibfnamefont {H.-W.}\ \bibnamefont {{Rix}}}, \
  and\ \bibinfo {author} {\bibfnamefont {M.~K.}\ \bibnamefont {{Ness}}},\
  }\href {\doibase 10.3847/1538-4357/aaf648} {\bibfield  {journal} {\bibinfo
  {journal} {\apj}\ }\textbf {\bibinfo {volume} {871}},\ \bibinfo {eid} {120}
  (\bibinfo {year} {2019})},\ \Eprint {http://arxiv.org/abs/1810.09466}
  {arXiv:1810.09466 [astro-ph.GA]} \BibitemShut {NoStop}%
\bibitem [{\citenamefont {Abuter}\ \emph {et~al.}(2018)\citenamefont {Abuter}
  \emph {et~al.}}]{Abuter:2018drb}%
  \BibitemOpen
  \bibfield  {author} {\bibinfo {author} {\bibfnamefont {R.}~\bibnamefont
  {Abuter}} \emph {et~al.} (\bibinfo {collaboration} {GRAVITY}),\ }\href
  {\doibase 10.1051/0004-6361/201833718} {\bibfield  {journal} {\bibinfo
  {journal} {Astron. Astrophys.}\ }\textbf {\bibinfo {volume} {615}},\ \bibinfo
  {pages} {L15} (\bibinfo {year} {2018})},\ \Eprint
  {http://arxiv.org/abs/1807.09409} {arXiv:1807.09409 [astro-ph.GA]}
  \BibitemShut {NoStop}%
\bibitem [{\citenamefont {{Bonaca}}\ \emph {et~al.}(2020)\citenamefont
  {{Bonaca}}, \citenamefont {{Conroy}}, \citenamefont {{Hogg}}, \citenamefont
  {{Cargile}}, \citenamefont {{Caldwell}}, \citenamefont {{Naidu}},
  \citenamefont {{Price-Whelan}}, \citenamefont {{Speagle}},\ and\
  \citenamefont {{Johnson}}}]{2020arXiv200107215B}%
  \BibitemOpen
  \bibfield  {author} {\bibinfo {author} {\bibfnamefont {A.}~\bibnamefont
  {{Bonaca}}}, \bibinfo {author} {\bibfnamefont {C.}~\bibnamefont {{Conroy}}},
  \bibinfo {author} {\bibfnamefont {D.~W.}\ \bibnamefont {{Hogg}}}, \bibinfo
  {author} {\bibfnamefont {P.~A.}\ \bibnamefont {{Cargile}}}, \bibinfo {author}
  {\bibfnamefont {N.}~\bibnamefont {{Caldwell}}}, \bibinfo {author}
  {\bibfnamefont {R.~P.}\ \bibnamefont {{Naidu}}}, \bibinfo {author}
  {\bibfnamefont {A.~M.}\ \bibnamefont {{Price-Whelan}}}, \bibinfo {author}
  {\bibfnamefont {J.~S.}\ \bibnamefont {{Speagle}}}, \ and\ \bibinfo {author}
  {\bibfnamefont {B.~D.}\ \bibnamefont {{Johnson}}},\ }\href@noop {} {\bibfield
   {journal} {\bibinfo  {journal} {arXiv e-prints}\ ,\ \bibinfo {eid}
  {arXiv:2001.07215}} (\bibinfo {year} {2020})},\ \Eprint
  {http://arxiv.org/abs/2001.07215} {arXiv:2001.07215 [astro-ph.GA]}
  \BibitemShut {NoStop}%
\bibitem [{\citenamefont {Garzilli}\ \emph {et~al.}(2019)\citenamefont
  {Garzilli}, \citenamefont {Ruchayskiy}, \citenamefont {Magalich},\ and\
  \citenamefont {Boyarsky}}]{Garzilli:2019qki}%
  \BibitemOpen
  \bibfield  {author} {\bibinfo {author} {\bibfnamefont {A.}~\bibnamefont
  {Garzilli}}, \bibinfo {author} {\bibfnamefont {O.}~\bibnamefont
  {Ruchayskiy}}, \bibinfo {author} {\bibfnamefont {A.}~\bibnamefont
  {Magalich}}, \ and\ \bibinfo {author} {\bibfnamefont {A.}~\bibnamefont
  {Boyarsky}},\ }\href@noop {} {\  (\bibinfo {year} {2019})},\ \Eprint
  {http://arxiv.org/abs/1912.09397} {arXiv:1912.09397 [astro-ph.CO]}
  \BibitemShut {NoStop}%
\bibitem [{\citenamefont {Archidiacono}\ \emph {et~al.}(2019)\citenamefont
  {Archidiacono}, \citenamefont {Hooper}, \citenamefont {Murgia}, \citenamefont
  {Bohr}, \citenamefont {Lesgourgues},\ and\ \citenamefont
  {Viel}}]{Archidiacono:2019wdp}%
  \BibitemOpen
  \bibfield  {author} {\bibinfo {author} {\bibfnamefont {M.}~\bibnamefont
  {Archidiacono}}, \bibinfo {author} {\bibfnamefont {D.~C.}\ \bibnamefont
  {Hooper}}, \bibinfo {author} {\bibfnamefont {R.}~\bibnamefont {Murgia}},
  \bibinfo {author} {\bibfnamefont {S.}~\bibnamefont {Bohr}}, \bibinfo {author}
  {\bibfnamefont {J.}~\bibnamefont {Lesgourgues}}, \ and\ \bibinfo {author}
  {\bibfnamefont {M.}~\bibnamefont {Viel}},\ }\href {\doibase
  10.1088/1475-7516/2019/10/055} {\bibfield  {journal} {\bibinfo  {journal}
  {JCAP}\ }\textbf {\bibinfo {volume} {1910}},\ \bibinfo {pages} {055}
  (\bibinfo {year} {2019})},\ \Eprint {http://arxiv.org/abs/1907.01496}
  {arXiv:1907.01496 [astro-ph.CO]} \BibitemShut {NoStop}%
\bibitem [{\citenamefont {Bonaca}\ \emph {et~al.}(2018)\citenamefont {Bonaca},
  \citenamefont {Hogg}, \citenamefont {Price-Whelan},\ and\ \citenamefont
  {Conroy}}]{Bonaca:2018fek}%
  \BibitemOpen
  \bibfield  {author} {\bibinfo {author} {\bibfnamefont {A.}~\bibnamefont
  {Bonaca}}, \bibinfo {author} {\bibfnamefont {D.~W.}\ \bibnamefont {Hogg}},
  \bibinfo {author} {\bibfnamefont {A.~M.}\ \bibnamefont {Price-Whelan}}, \
  and\ \bibinfo {author} {\bibfnamefont {C.}~\bibnamefont {Conroy}},\ }\href
  {\doibase 10.3847/1538-4357/ab2873} {\  (\bibinfo {year} {2018}),\
  10.3847/1538-4357/ab2873},\ \Eprint {http://arxiv.org/abs/1811.03631}
  {arXiv:1811.03631 [astro-ph.GA]} \BibitemShut {NoStop}%
\bibitem [{\citenamefont {{Mateu}}\ \emph {et~al.}(2018)\citenamefont
  {{Mateu}}, \citenamefont {{Read}},\ and\ \citenamefont
  {{Kawata}}}]{2018MNRAS.474.4112M}%
  \BibitemOpen
  \bibfield  {author} {\bibinfo {author} {\bibfnamefont {C.}~\bibnamefont
  {{Mateu}}}, \bibinfo {author} {\bibfnamefont {J.~I.}\ \bibnamefont {{Read}}},
  \ and\ \bibinfo {author} {\bibfnamefont {D.}~\bibnamefont {{Kawata}}},\
  }\href {\doibase 10.1093/mnras/stx2937} {\bibfield  {journal} {\bibinfo
  {journal} {\mnras}\ }\textbf {\bibinfo {volume} {474}},\ \bibinfo {pages}
  {4112} (\bibinfo {year} {2018})},\ \Eprint {http://arxiv.org/abs/1711.03967}
  {arXiv:1711.03967 [astro-ph.GA]} \BibitemShut {NoStop}%
\bibitem [{\citenamefont {Morinaga}\ \emph {et~al.}(2019)\citenamefont
  {Morinaga}, \citenamefont {Ishiyama}, \citenamefont {Kirihara},\ and\
  \citenamefont {Kinjo}}]{Morinaga:2019fsy}%
  \BibitemOpen
  \bibfield  {author} {\bibinfo {author} {\bibfnamefont {Y.}~\bibnamefont
  {Morinaga}}, \bibinfo {author} {\bibfnamefont {T.}~\bibnamefont {Ishiyama}},
  \bibinfo {author} {\bibfnamefont {T.}~\bibnamefont {Kirihara}}, \ and\
  \bibinfo {author} {\bibfnamefont {K.}~\bibnamefont {Kinjo}},\ }\href
  {\doibase 10.1093/mnras/stz1373} {\bibfield  {journal} {\bibinfo  {journal}
  {Mon. Not. Roy. Astron. Soc.}\ }\textbf {\bibinfo {volume} {487}},\ \bibinfo
  {pages} {2718} (\bibinfo {year} {2019})},\ \Eprint
  {http://arxiv.org/abs/1901.04748} {arXiv:1901.04748 [astro-ph.GA]}
  \BibitemShut {NoStop}%
\bibitem [{\citenamefont {{Molin{\'e}}}\ \emph {et~al.}(2017)\citenamefont
  {{Molin{\'e}}}, \citenamefont {{S{\'a}nchez-Conde}}, \citenamefont
  {{Palomares-Ruiz}},\ and\ \citenamefont {{Prada}}}]{2017MNRAS.466.4974M}%
  \BibitemOpen
  \bibfield  {author} {\bibinfo {author} {\bibfnamefont {{\'A}.}~\bibnamefont
  {{Molin{\'e}}}}, \bibinfo {author} {\bibfnamefont {M.~A.}\ \bibnamefont
  {{S{\'a}nchez-Conde}}}, \bibinfo {author} {\bibfnamefont {S.}~\bibnamefont
  {{Palomares-Ruiz}}}, \ and\ \bibinfo {author} {\bibfnamefont
  {F.}~\bibnamefont {{Prada}}},\ }\href {\doibase 10.1093/mnras/stx026}
  {\bibfield  {journal} {\bibinfo  {journal} {\mnras}\ }\textbf {\bibinfo
  {volume} {466}},\ \bibinfo {pages} {4974} (\bibinfo {year} {2017})},\ \Eprint
  {http://arxiv.org/abs/1603.04057} {arXiv:1603.04057 [astro-ph.CO]}
  \BibitemShut {NoStop}%
\bibitem [{\citenamefont {{Calore}}\ \emph {et~al.}(2019)\citenamefont
  {{Calore}}, \citenamefont {{H{\"u}tten}},\ and\ \citenamefont
  {{Stref}}}]{2019Galax...7...90C}%
  \BibitemOpen
  \bibfield  {author} {\bibinfo {author} {\bibfnamefont {F.}~\bibnamefont
  {{Calore}}}, \bibinfo {author} {\bibfnamefont {M.}~\bibnamefont
  {{H{\"u}tten}}}, \ and\ \bibinfo {author} {\bibfnamefont {M.}~\bibnamefont
  {{Stref}}},\ }\href {\doibase 10.3390/galaxies7040090} {\bibfield  {journal}
  {\bibinfo  {journal} {Galaxies}\ }\textbf {\bibinfo {volume} {7}},\ \bibinfo
  {pages} {90} (\bibinfo {year} {2019})},\ \Eprint
  {http://arxiv.org/abs/1910.13722} {arXiv:1910.13722 [astro-ph.HE]}
  \BibitemShut {NoStop}%
\bibitem [{\citenamefont {Schutz}(2020)}]{Schutz:2020jox}%
  \BibitemOpen
  \bibfield  {author} {\bibinfo {author} {\bibfnamefont {K.}~\bibnamefont
  {Schutz}},\ }\href@noop {} {\  (\bibinfo {year} {2020})},\ \Eprint
  {http://arxiv.org/abs/2001.05503} {arXiv:2001.05503 [astro-ph.CO]}
  \BibitemShut {NoStop}%
\bibitem [{\citenamefont {Lewis}\ \emph {et~al.}(2000)\citenamefont {Lewis},
  \citenamefont {Challinor},\ and\ \citenamefont {Lasenby}}]{Lewis:1999bs}%
  \BibitemOpen
  \bibfield  {author} {\bibinfo {author} {\bibfnamefont {A.}~\bibnamefont
  {Lewis}}, \bibinfo {author} {\bibfnamefont {A.}~\bibnamefont {Challinor}}, \
  and\ \bibinfo {author} {\bibfnamefont {A.}~\bibnamefont {Lasenby}},\ }\href
  {\doibase 10.1086/309179} {\bibfield  {journal} {\bibinfo  {journal}
  {Astrophys. J.}\ }\textbf {\bibinfo {volume} {538}},\ \bibinfo {pages} {473}
  (\bibinfo {year} {2000})},\ \Eprint {http://arxiv.org/abs/astro-ph/9911177}
  {arXiv:astro-ph/9911177 [astro-ph]} \BibitemShut {NoStop}%
\bibitem [{\citenamefont {Hlozek}\ \emph {et~al.}(2015)\citenamefont {Hlozek},
  \citenamefont {Grin}, \citenamefont {Marsh},\ and\ \citenamefont
  {Ferreira}}]{Hlozek:2014lca}%
  \BibitemOpen
  \bibfield  {author} {\bibinfo {author} {\bibfnamefont {R.}~\bibnamefont
  {Hlozek}}, \bibinfo {author} {\bibfnamefont {D.}~\bibnamefont {Grin}},
  \bibinfo {author} {\bibfnamefont {D.~J.~E.}\ \bibnamefont {Marsh}}, \ and\
  \bibinfo {author} {\bibfnamefont {P.~G.}\ \bibnamefont {Ferreira}},\ }\href
  {\doibase 10.1103/PhysRevD.91.103512} {\bibfield  {journal} {\bibinfo
  {journal} {Phys. Rev.}\ }\textbf {\bibinfo {volume} {D91}},\ \bibinfo {pages}
  {103512} (\bibinfo {year} {2015})},\ \Eprint {http://arxiv.org/abs/1410.2896}
  {arXiv:1410.2896 [astro-ph.CO]} \BibitemShut {NoStop}%
\bibitem [{\citenamefont {Du}(2018)}]{Du:2018wxl}%
  \BibitemOpen
  \bibfield  {author} {\bibinfo {author} {\bibfnamefont {X.~L.}\ \bibnamefont
  {Du}},\ }\emph {\bibinfo {title} {{Structure Formation with Ultralight Axion
  Dark Matter}}},\ \href@noop {} {Ph.D. thesis},\ \bibinfo  {school} {Gottingen
  U.} (\bibinfo {year} {2018})\BibitemShut {NoStop}%
\bibitem [{\citenamefont {Lovell}(2020)}]{Lovell:2020bcy}%
  \BibitemOpen
  \bibfield  {author} {\bibinfo {author} {\bibfnamefont {M.~R.}\ \bibnamefont
  {Lovell}},\ }\href@noop {} {\  (\bibinfo {year} {2020})},\ \Eprint
  {http://arxiv.org/abs/2003.01125} {arXiv:2003.01125 [astro-ph.CO]}
  \BibitemShut {NoStop}%
\end{thebibliography}%

\end{document}